\documentclass[12pt]{article}
\usepackage{amsmath,amssymb,amsfonts}

\usepackage[nosort]{cite}

\usepackage{hyperref}
\usepackage{young}
\usepackage[vcentermath]{youngtab}
\usepackage{tikz}

\usepackage{color,xcolor}
\definecolor{darkred}{rgb}{0.65,0.15,0}
\definecolor{newgreen}{rgb}{0.2,0.62,0.14}
\hypersetup{pdfborder={0 0 0},colorlinks=true,urlcolor=blue,citecolor=blue,linkcolor=darkred,linktocpage=true}

\numberwithin{equation}{section} 

\newcommand{\ALT}{\textrm{\large{$\wedge$}}}

\newcommand{\nn}{\nonumber}

\begin{document}

 \mbox{}
\vspace{20mm}

\begin{center}
{\bf \Large Infinite-dimensional algebras as extensions\\[4mm] of kinematic algebras}\\[12mm]

  Joaquim Gomis${}^1$ and Axel Kleinschmidt${}^{2,3}$\\[3mm]
\footnotemark[1]{\it Departament de F\'isica Qu\`antica i Astrof\'isica\\
 and Institut de Ci\`encies del Cosmos (ICCUB), Universitat de Barcelona\\ 
Mart\'i i Franqu\`es , ES-08028 Barcelona, Spain}\\[1mm]
\footnotemark[2]{\it  Max-Planck-Institut f\"ur Gravitationsphysik\\
     Albert-Einstein-Institut \\
     Am M\"uhlenberg 1, D-14476 Potsdam, Germany}\\[1mm]
\footnotemark[3]{\it International Solvay Institutes\\
ULB-Campus Plaine CP231, BE-1050 Brussels, Belgium}\\[20mm]

\begin{abstract}

Kinematic algebras can be realised on geometric spaces and constrain the physical models that can live on these spaces. Different types of kinematic algebras exist and we consider the interplay of these algebras for non-relativistic limits of a relativistic system, including both the Galilei and the Carroll limit. 
We develop a framework that captures systematically the corrections to the strict non-relativistic limit by introducing new infinite-dimensional  algebras, with emphasis on the Carroll case. 
One of our results is to highlight a new type of duality between Galilei and Carroll limits that extends to corrections as well.
We realise these algebras in terms of particle models.
Other applications include curvature corrections and particles in a background electro-magnetic field.

\end{abstract}
\end{center}

\thispagestyle{empty}

\newpage
\setcounter{page}{1}

\setcounter{tocdepth}{2}
\tableofcontents

\bigskip
\hrule
\medskip

\section{Introduction}

Relativistic and non-relativistic systems are usually distinguished by their kinematic algebras. 
Classifications of possible kinematic algebras (for point particles) have been obtained in four space-time dimensions in~\cite{Bacry:1968zf,Bacry:1986} and in  different dimensions for instance in~\cite{Figueroa-OFarrill:2017ycu,Figueroa-OFarrill:2018ilb}.
The Poincar\'e algebra constrains relativistic systems in flat space, while non-relativistic systems of particles are subjected for example to Galilean or Carrollian algebras.\footnote{In this paper we use `non-relativistic' generally to refer to any system that does not have Poincar\'e symmetry.}   In a given system, there is typically a physical quantity that can be combined with the speed of light $c$ to form a dimensionless quantity whose limit to zero (or to infinity) defines the non-relativistic limit of the system. The prototypical example is the speed of a particle $v$ and then the dimensionless quantity can be taken as $v/c$. In gravitational systems there is additionally the Newton constant to form dimensionless parameters and multiple limits can be considered in such cases~\cite{Buonanno:1998gg,Blanchet:2013haa}. Similarly, for extended objects one can consider non-relativistic limits, but the extended nature of the world-volume allows for a variety of different limits~\cite{Brugues:2004an,Andringa:2012uz,Batlle:2016iel,Barducci:2019jhj}.

The kinematic algebras do not contain the dimensionful quantity of a given physical system but only fundamental constants, for example $c$ in the case of the Poincar\'e algebra although it is usually not explicit since it is absorbed in the definition of the generators. One can still consider non-relativistic limits of the algebras by formally scaling the parameters. This process, first performed for obtaining the Galilei algebra from the Poincar\'e algebra, is known as a (In\"on\"u--Wigner) contraction of a Lie algebra~\cite{Inonu:1953sp}. The limit $c\to\infty$ considered there implicitly assumes that \textit{all} particle velocities that can arise are small compared to the speed of light. The similar contraction $c\to 0$ giving the Carroll algebra~\cite{LevyLeblond:1965,gupta1966analogue} makes the  assumption that all particle velocities are large compared to the speed of light. We shall review these cases in detail in section~\ref{sec:contr}. 

The non-relativistic symmetries obtained by In\"on\"u--Wigner contractions are the strict limits of the scaling parameter, e.g. $c\to \infty$ or $c\to 0$.  However, the inclusion of relativistic corrections is desirable for many systems such as their contributions to the fine structure of atoms. While many relativistic equations can be expanded in the small parameters, the resulting equations do not exhibit symmetries beyond the contracted algebra. 
It is one of the aims of this contribution to explain how to formulate a framework of kinematic algebras for these corrections. The main method we will use is to enlarge the space on which the symmetry acts and thereby allow for extensions of the contracted algebra.

One can construct systematically kinematic algebras that include perturbative corrections in a parameter (like $1/c$) by deforming 
the contracted algebra, in this way performing something like the inverse of a Lie algebra contraction. This process adds new generators to the original algebra and thus enlarges it. 
Considering all-order expansions in the small parameter we obtain infinite-dimensional algebras and there are different ways of arriving at them. 

One is to view the deformation problem as a cohomology problem such that one is asking for the most general non-trivial commutator of generators that commute in the contracted algebra. 
For instance, in the Galilei algebra one has famously that spatial translations $P_i$ and Galilean boosts $B_j$ commute while their most general commutator would be $[P_i,B_j]=Z_{i,j}+\tfrac12 B_{ij}$, containing both a symmetric ($Z_{i,j}=Z_{j,i})$ and an anti-symmetric ($B_{ij}=-B_{ji}$) part and all these generators can be considered as new.\footnote{The trace $\delta^{ij} Z_{i,j}$ is proportional to the Bargmann extension associated with massive representations of the Galilei algebra. The general symmetric $Z_{i,j}$ can be thought of as the anisotropic mass $m_{ij}$ of a particle~\cite{Bogoslovsky:1983,Bonanos:2008kr}.}
Adding these generators to the original algebra one can repeat the question of what is now the most general commutator of the elements of this new algebra, for instance of $P_i$ with $B_{jk}$, and introduce new elements in this way. The algebra obtained in this way typically involves a \textit{free} Lie algebra where all possible commutators are present; in this case the Galilean free Lie algebra of~\cite{Gomis:2019fdh}.\footnote{In our construction, it will be the semi-direct product of a manifest covariance algebra (e.g. spatial rotations) with a free Lie algebra. This is why we  only say `includes'.}
Free Lie algebras are characterised by containing all possible commutators of a basic set of elements such that the Jacobi identity is satisfied. They are infinite-dimensional and possess many interesting quotient Lie algebras whose physical interpretation can differ as we shall show in many examples.

A second way of obtaining a systematic perturbative extension of a kinematic algebra is the method of \textit{Lie algebra expansion}~\cite{Hatsuda:2001pp,Boulanger:2002bt,deAzcarraga:2002xi,Izaurieta:2006zz,deAzcarraga:2007et,Bergshoeff:2019ctr}. This method can be described in its simplest form by starting with an algebra and tensoring it with a commutative semigroup. An example of a commutative semigroup is given by $S_E^{(N)} = \{ \lambda^n \,|\, 0\leq n \leq N\} \cup \{0\}$ such that $\lambda^n \lambda^m =\lambda^{m+n}$ for $m+n\leq N$ and the product is zero otherwise. Clearly, there is a relation of this structure to the expansion up to order $N$ in a small parameter $\lambda$, thinking of $0$ as $\mathcal{O}(\lambda^{N+1})$. Typically, one uses refinements of this construction that we shall discuss in more detail in the main body of the paper and a full perturbative treatment including all terms up to $N\to\infty$ then relies on formal power series in $\lambda$.

The two constructions are not unrelated. This can be seen in the Galilean example with free Lie algebra commutator $[P_i,B_j]=Z_{i,j}+\tfrac12 B_{ij}$ as follows. Maintaining spatial rotation invariance, we can separate out the trace of the symmetric generator $Z=\delta^{ij} Z_{i,j}$ and from the structure of the Poincar\'e algebra we would expect this term to be related to temporal relativistic translation $P_0$. However, it is more useful to consider $Z$ as the first relativistic correction to the non-relativistic Hamiltonian and thus as something of order $c^{-2}$.
The relation to a Lie algebra expansion is thus achieved by considering $Z$ as $P_0\otimes c^{-2}$ and $P_i$ and $B_i$ as $P_i\otimes c^{-1}$ and $M_{i0}\otimes c^{-1}$. If one quotients the free Lie algebra by the tracefree part of $Z_{i,j}$ and the anti-symmetric $B_{ij}$, the commutator $[P_i,B_j] =\delta_{ij} Z$ agrees exactly with that of the expanded algebra to order $c^{-2}$.\footnote{The algebra with $[P_i,B_j]=\delta_{ij}Z$ is known as the Bargmann central extension of the Galilei algebra~\cite{Bargmann:1954gh}.} Therefore the Lie algebra expansion is a special case of a free Lie algebra construction.
The Carrollian limit is obtained by exchanging
$P_0\leftrightarrow P_i$ \cite{Barducci:2018wuj,Barducci:2019jhj}, see the next section for more details. 
We generalise this duality to the infinite-dimensional case where corrections are included in the kinematic algebra.

There is furthermore a connection of the two constructions to (Borel subalgebras of) affine Kac--Moody algebras. Affine Kac--Moody algebras are obtained by tensoring a finite-dimensional algebra $\mathfrak{g}$ with Laurent polynomial rings in a variable, say $\lambda$, possibly additionally twisted by an outer automorphism. 
Restricting to ordinary polynomials rather than Laurent polynomials one describes the subalgebra of non-negative levels of the Kac--Moody algebra and clearly this agrees with a specific Lie algebra expansion of $\mathfrak{g}$. At the same time, the subalgebra of positive levels is the quotient algebra of a free Lie algebra~\cite{Kac:1990}, so that also the viewpoint of the free Lie algebra enters in this relation. We shall describe this connection in more detail.

There are many variations of these constructions one can consider, depending on the algebra one starts with and the precise expansion or quotient of a free Lie algebra one takes. The physical interpretation of the parameter $\lambda$ depends on the context one considers and is by no means restricted to non-relativistic limits. It can also be viewed as a curvature parameter when one wants to describe deviations from flat space isometries as we shall review. Another arena is where higher powers of $\lambda$ correspond to more and more complicated electro-magnetic backgrounds in Minkowski space, where the kinematic algebra becomes a generalisation of the Maxwell algebra~\cite{Bacry:1970ye,Schrader:1972zd,Bonanos:2008ez,Gomis:2017cmt}. 

The above methods provide a plethora of kinematic algebras of finite or even infinite dimension. Our next aim will be to describe a space on which they can act in the same way that the Poincar\'e algebra acts on flat Minkowski space. Such a space is not hard to find using a non-linear realisation~\cite{Coleman:1969sm,Callan:1969sn,Salam:1969rq,Isham:1971dv,Volkov:1973vd,Ogievetsky:1974} approach with a suitable coset. 
It is of higher dimension than usual space-time and we present many examples with different physical interpretations. 
Once a generalised space is defined we strive to probe it using a physical model. The simplest instance is that of a free particle moving in it and there are canonical constructions of associated particle models that we shall go through.\footnote{In the case of gravity, $1/c^2$ corrections have been considered already for example in~\cite{Dautcourt:1990,VandenBleeken:2017rij}.} One could also consider using higher-dimensional objects as probes but we shall not pursue this here.

As we shall demonstrate, these particle models give rise systematically to relativistic (or similar) corrections to the dynamics associated with the truncated algebra. Particular emphasis will be put on the case of Carroll particles, both of ordinary 
\cite{Bergshoeff:2014jla,Duval:2014uoa}
and of tachyonic~\cite{deBoer:2021jej} type. The reason for this is that they have featured prominently in recent studies, including applications to cosmology~\cite{Bergshoeff:2017btm,Gomis:2019nih,Henneaux:2021yzg,deBoer:2021jej,Hansen:2021fxi}. Formally, the Carroll limit is also related to the Belinsky--Khalatnikov--Lifshitz limit~\cite{Belinsky:1970ew,Damour:2002et} where temporal derivatives dominate over spatial derivatives $\partial_t \gg c \partial_x$ and so formally $c \to 0$. 
We shall also exhibit a new type of duality between (corrections to) Galilei and Carroll particle actions in section~\ref{sec:GalCar}.

The structure of this contribution is as follows. We first explain the basic algebraic constructions of kinematic algebras and their interrelations in section~\ref{sec:alg}. Then we present generalised space-times on which the kinematic algebras can act in section~\ref{sec:geo}. To probe the set-up we then consider free particle actions in section~\ref{sec:act} where we deduce non-relativistic and similar corrections. Some concluding comments are given in section~\ref{sec:concl}.


\section{Algebraic constructions}
\label{sec:alg}

We present various methods for constructing kinematic algebras and how they are related to one another.

\subsection{Contractions and extensions of kinematic algebras}
\label{sec:contr}

As an illustrative starting point we choose the Poincar\'e algebra in $D=d+1$ space-time dimensions
\begin{align}
\label{eq:Poin}
[ M_{ab}, M_{cd} ] &= \eta_{bc} M_{ad} - \eta_{ac} M_{bd} -\eta_{bd} M_{ac} + \eta_{ad} M_{bc}\,,\nn\\
{}[ M_{ab}, P_c ] &= \eta_{bc} P_a - \eta_{ac} P_b \,,\\
{}[ P_a, P_b ] &= 0\,,\nn
\end{align}
where small Roman indices from the beginning of the alphabet are fundamental $\mathfrak{so}(1,d)$ indices, e.g. $a=0,1,\ldots,d$, and $\eta_{ab}$ is the flat Minkowski metric of signature $(-++\ldots +)$.
When separating the time and space indices according to $a=(0,i)$ with $i=1,\ldots,d$, we let~\cite{LevyLeblond:1965}
\begin{align}
\label{eq:Gcontrlevy}
J_{ij} &= M_{ij}\,, &B_i &= \lambda^{1/2} M_{i0}\,,\nn\\
T_i &= \lambda^{1/2}P_i\,, & H &= P_0 
\end{align}
which is an invertible change of basis for any $\lambda>0$ and the algebra becomes
\begin{align}
[ J_{ij} , J_{kl} ] &= \delta_{jk} J_{il} - \delta_{ij} J_{ik} - \delta_{jl} J_{ik} + \delta_{il} J_{jk} \,,\nn\\
[ J_{ij} , B_k ] &= \delta_{jk} B_i - \delta_{ik} B_j\,,\nn\\ 
[ J_{ij} , T_k ] &=  \delta_{jk} T_i - \delta_{ik} T_j\,,\nn\\
[ J_{ij} , H ] &=0\,,\nn\\ 
[ B_i , B_j ] &= \lambda J_{ij} \,,\nn\\
[ B_i , T_j ] &= - \lambda \delta_{ij} H\,,\nn\\
[ B_i, H ] &= - T_i\,,\nn\\
[ T_i, T_j ] &= [T_i, H] =0\,.
\end{align}
We see that we can take the limit $\lambda\to 0$ smoothly and obtain a new algebra in that limit. This so-called \textit{contracted} algebra is the non-relativistic Galilei algebra ($c\to\infty$) where now Galilean boosts commute among themselves and with translations. 
This is the most famous example of an In\"on\"u--Wigner contraction of a Lie algebra~\cite{Inonu:1953sp}. As is usual, the contracted algebra is no longer isomorphic to the algebra with $\lambda>0$. The square root in~\eqref{eq:Gcontrlevy} arises since we think of $\lambda$ as $1/c^2$.\footnote{The same contraction can also be achieved when replacing the second line of~\eqref{eq:Gcontrlevy} by $T_i=P_i$ and $H = \lambda^{-1/2} P_0$. The two choices are related by an overall scaling of the translation generators $P_a \leftrightarrow \lambda^{1/2} P_a$ in the Poincar\'e algebra which, for $\lambda > 0$ is an invertible basis redefinition. For $\lambda^{1/2}=c^{-1}$ this includes changing the dimension of the translation generators.}
For future reference we write the resulting contracted Galilei algebra
\begin{align}
\label{eq:GalAlg}
[ J_{ij} , J_{kl} ] &= \delta_{jk} J_{il} - \delta_{ij} J_{ik} - \delta_{jl} J_{ik} + \delta_{il} J_{jk} \,,\nn\\
[ J_{ij} , B_k ] &= \delta_{jk} B_i - \delta_{ik} B_j\,,\nn\\ 
[ J_{ij} , T_k ] &=  \delta_{jk} T_i - \delta_{ik} T_j\,,\nn\\
[ J_{ij} , H ] &=0\,,\nn\\ 
[ B_i , B_j ] &=0 \,,\nn\\
[ B_i , T_j ] &= 0\,,\nn\\
[ B_i, H ] &= - T_i\,,\nn\\
[ T_i, T_j ] &= [T_i, H] =0\,.
\end{align}

An alternative contraction of the algebra is obtained by formally interchanging the roles of time and space directions for the translation generators, i.e., letting~\cite{LevyLeblond:1965}
\begin{align}
\label{eq:Ccontr}
J_{ij} &= M_{ij}\,,& K_i &= \lambda^{1/2} M_{i0}\,,\nn\\
T_i &= P_i\,,& K &= \lambda^{1/2} P_0 
\end{align}
and contracting again $\lambda\to 0$. This leads to the Carroll algebra
\begin{align}
\label{eq:Calg}
[ J_{ij} , J_{kl} ] &= \delta_{jk} J_{il} - \delta_{ij} J_{ik} - \delta_{jl} J_{ik} + \delta_{il} J_{jk} \,,\nn\\
[ J_{ij} , K_k ] &= \delta_{jk} K_i - \delta_{ik} K_j\,,\nn\\ 
[ J_{ij} , T_k ] &=  \delta_{jk} T_i - \delta_{ik} T_j\,,\nn\\
[ J_{ij} , K ] &=0\,,\nn\\ 
[ K_i , K_j ] &= 0 \,,\nn\\
[ K_i , T_j ] &= -  \delta_{ij} K\,,\nn\\
[ K_i, K ] &= 0\,,\nn\\
[ T_i, T_j ] &= [T_i, K] =0\,.
\end{align}
This contraction of the Poincar\'e algebra is also known as the Carroll limit in which the speed of light tends to zero ($c\to 0$).
We see from~\eqref{eq:Gcontrlevy} and~\eqref{eq:Ccontr} that there is a duality between the Galilean and the Carrollian contraction that simply exchanges the role of space and time translations in the contraction. Thinking of the time direction as the longitudinal direction of the world-line of a (massive) particle and the space directions as the transverse directions, makes it clear that a similar duality between longitudinal and transverse directions will be present for contractions related to extended objects as was discussed in more detail in~\cite{Barducci:2019jhj}.

Let us now formalise the contraction process. We start from an algebra $\mathfrak{g}$ with generators $t_\alpha$ and structures constants $f_{\alpha\beta}{}^\gamma$. Then for each $\lambda> 0$ we define a Lie algebra isomorphism $c_\lambda : \mathfrak{g} \to \mathfrak{g}_\lambda$ to another algebra $\mathfrak{g}_\lambda$. If the limit $\lambda\to 0$ is well-defined, we call the limiting algebra $\mathfrak{g}_0$ the contracted algebra. Note that at $\lambda=0$ we no longer necessarily have a Lie algebra isomorphism. In the examples above, the isomorphisms for $\lambda>0$ were given in~\eqref{eq:Gcontrlevy} and~\eqref{eq:Ccontr}, respectively. 
Contractions preserve the number of generators but the resulting algebra is not necessarily isomorphic to the starting one. Moreover, it is not generally possible to reverse the contraction process directly.

One approach to undoing the contraction perturbatively requires the knowledge of the original algebra $\mathfrak{g}$. Writing its generators multiplied by a formal power series in $\lambda$
\begin{align}
\label{eq:tan}
t_\alpha \to \sum_{n\geq 0} t_\alpha \otimes \lambda^{n_0(\alpha) + n} = \sum_{n\geq 0} t_\alpha^{(n)} \,,
\end{align}
where the offset $n_0(\alpha)$ can depend on the generator, we construct an infinite-dimensional algebra of the generators $t_\alpha^{(n)}$. 
We refer to the generator $t_\alpha^{(n)}$ as `level $n$' and think of it as the $n$th order perturbative expansion in the parameter $\lambda$. The offset should be chosen in such a way that the commutators of level $m$ with level $n$ only contain generators of level $\geq m{+}n$, where the commutator is defined in using the associative product of power series together with the Lie bracket on $\mathfrak{g}$.
The lowest order commutators involving only the $t_\alpha^{(0)}$ then correspond to the contracted algebra, but the higher terms capture the perturbative expansion of the original of the original algebra $\mathfrak{g}$. 
We denote by $\mathfrak{g}^{(N)}$ the algebra obtained by keeping terms up to level $N$. In this way, $\mathfrak{g}^{(0)} = \mathfrak{g}_0$, the contraction of $\mathfrak{g}$. When we keep all levels, we shall use the notation $\mathfrak{g}^{(\infty)}$.\footnote{In the construction, we are assuming for simplicity that we have a basis of $\mathfrak{g}$ in which the contraction works by simply rescaling the basis generators. There are also contractions where this assumption is not satisfied, see for example the contraction of Poincar\'e$\,\oplus\, \mathfrak{gl}(1)$ to the Bargmann algebra in~\cite{Aldaya:1985plo}.}

Let us exemplify this in the case of the Galilei algebra~\eqref{eq:GalAlg}. We define for $n\geq 0$
\begin{align}
\label{eq:galexp}
J_{ij}^{(n)} &= M_{ij} \otimes \lambda^n\,,& B_i^{(n)} &= M_{i0} \otimes \lambda^{1/2+n}\,,\nn\\
T_i^{(n)} &= P_i \otimes \lambda^{1/2+n} \,,& H^{(n)} &= P_0 \otimes \lambda^{n}\,,
\end{align}
where the offsets are taken in accordance with~\eqref{eq:Gcontrlevy}.
The associated Lie algebra is
\begin{align}
\label{eq:infG}
[ J_{ij}^{(m)} , J_{kl}^{(n)} ] &= \delta_{jk} J_{il}^{(m+n)} - \delta_{ij} J_{ik}^{(m+n)} - \delta_{jl} J_{ik}^{(m+n)} + \delta_{il} J_{jk}^{(m+n)} \,,\nn\\
[ J_{ij}^{(m)} , B_k^{(n)} ] &= \delta_{jk} B_i^{(m+n)} - \delta_{ik} B_j^{(m+n)}\,,\nn\\ 
[ J_{ij}^{(m)} , T_k^{(n)} ] &=  \delta_{jk} T_i^{(m+n)} - \delta_{ik} T_j^{(m+n)}\,,\nn\\
[ J_{ij}^{(m)} , H^{(n)} ] &=0\,,\nn\\ 
[ B_i^{(m)} , B_j ^{(n)}] &= J_{ij}^{(m+n+1)} \,,\nn\\
[ B_i^{(m)} , T_j^{(n)} ] &= - \delta_{ij} H^{(m+n+1)}\,,\nn\\
[ B_i^{(m)}, H^{(n)} ] &= - T_i^{(m+n)}\,,\nn\\
[ T_i^{(m)}, T_j^{(n)} ] &= [T_i^{(m)}, H^{(n)}] =0\,.
\end{align}
Since all commutators of generators at levels $m$ and $n$ generate only terms of level at least $m{+}n$, we can consistently quotient out all generators above a fixed level $N$. This leads to a finite-dimensional algebra. Retaining only the generators of level $0$ leads to the Galilei algebra that is obtained by contraction. 
Keeping all generators up to level $N$ then gives a perturbative approximation to the Poincar\'e algebra up to that order. The algebra~\eqref{eq:infG} was given in~\cite{Gomis:2019sqv}, see 
also \cite{Khasanov:2011jr,Hansen:2019pkl,Ozdemir:2019orp,Hansen:2019vqf}.

Repeating the same construction for the Carroll contraction~\eqref{eq:Ccontr} one can start with
\begin{align}
\label{eq:carrollexp}
J_{ij}^{(n)} &= M_{ij} \otimes \lambda^n \,,& K_i^{(n)} &= M_{i0} \otimes \lambda^{1/2+n}\,,\nn\\
T_i^{(n)} &= P_i\otimes \lambda^n\,,& K^{(n)} &= P_0 \otimes \lambda^{1/2+n}
\end{align}
We note that, when comparing~\eqref{eq:galexp} for the expanded  Galilei algebra with~\eqref{eq:carrollexp} for the Carroll algebra, there is a duality between the two algebras where the $\lambda^{1/2}$ is changed from $P_0$ to $P_i$. This is a generalisation of the type of duality that has been noted before in~\cite{Barducci:2018wuj,Barducci:2019jhj,Bergshoeff:2020xhv}.

The definition~\eqref{eq:carrollexp} leads to the infinite-dimensional algebra 
\begin{align}
\label{eq:infC}
[ J_{ij}^{(m)} , J_{kl}^{(n)} ] &= \delta_{jk} J_{il}^{(m+n)} - \delta_{ij} J_{ik}^{(m+n)} - \delta_{jl} J_{ik}^{(m+n)} + \delta_{il} J_{jk}^{(m+n)} \,,\nn\\
[ J_{ij}^{(m)} , K_k^{(n)} ] &= \delta_{jk} K_i^{(m+n)} - \delta_{ik} K_j^{(m+n)}\,,\nn\\ 
[ J_{ij}^{(m)} , T_k^{(n)} ] &=  \delta_{jk} T_i^{(m+n)} - \delta_{ik} T_j^{(m+n)}\,,\nn\\
[ J_{ij}^{(m)} , K^{(n)} ] &=0\,,\nn\\ 
[ K_i^{(m)} , K_j^{(n)} ] &= J_{ij}^{(m+n+1)} \,,\nn\\
[ K_i^{(m)} , T_j^{(n)} ] &= -  \delta_{ij} K^{(m+n)}\,,\nn\\
[ K_i^{(m)}, K^{(n)} ] &= -T_i^{(m+n+1)}\,,\nn\\
[ T_i^{(m)}, T_j^{(n)} ] &= [T_i^{(m)}, K^{(n)}] =0\,.
\end{align}
We note that comparing this formula to~\eqref{eq:infG}, there are subtle but important differences in the shifts of the indices by $+1$ on the right-hand sides which are due to the placements of $\lambda^{1/2}$ in the definitions of the algebra, and so ultimately to the physical meaning of the contractions.

The above procedure can also be viewed as a variant of the method of Lie algebra expansions that was originally introduced in~\cite{Hatsuda:2001pp,Boulanger:2002bt,deAzcarraga:2002xi,Izaurieta:2006zz,deAzcarraga:2007et,Bergshoeff:2019ctr}. For a Lie algebra expansion in its formulation given in~\cite{Izaurieta:2006zz} one requires an abelian semi-group $S$ whose elements we call $\lambda_i$ and the S-expanded Lie algebra $\mathfrak{g}\times S$ has a basis $t_\alpha \otimes \lambda_i$ and the Lie bracket
\begin{align}
\left[ t_\alpha\otimes \lambda_i , t_\beta \otimes \lambda_j \right] = f_{\alpha\beta}{}^\gamma t_\gamma\otimes \lambda_i\lambda_j
\end{align}
and commutativity of the product on $S$ ensures the Jacobi identity of the expanded algebra. 
A simple example of a semi-group is given by $S_E^{(N)}= \{\lambda_0,\ldots,\lambda_{N},\lambda_{N+1}\}$ with abelian product
\begin{align}
\lambda_i\lambda_j = \left\{ \begin{array}{cl} 
\lambda_{i+j}  & \text{if $i+j\leq N$}\\
\lambda_{N+1} & \text{otherwise}
\end{array}\right.
\end{align}
The element $\lambda_{N+1}$ serves as a substitute for zero in the multiplication.
Tensoring this semi-group with the real numbers corresponds to taking the quotient of the polynomial rings $\mathbb{R}[\lambda] / (\lambda^{N+1} \mathbb{R}[\lambda])$, i.e. working perturbatively in $\lambda$ up to order $N$. In this identification we have simply $\lambda_i = \lambda^i$, i.e., the $i$th basis of the semi-group should be identified with the $i$th power of the expansion parameter $\lambda$. One can also take the limit $N\to\infty$ and work with formal power series.

A more refined version of the Lie algebra expansion method can be obtained when the original Lie algebra has a decomposition. We here restrict to the case when\footnote{More general cases can be found in~\cite{Izaurieta:2006zz,deAzcarraga:2007et}.}
\begin{align}
\label{eq:expC}
\mathfrak{g} = V_0 \oplus V_1
\quad\text{with}\quad
[ V_0, V_0 ] \subset V_0\,,\quad [V_0, V_1] \subset V_1\,,\quad [V_1,V_1]\subset V_0\,.
\end{align}
(We use different letters here for the graded pieces in order to avoid confusion with the contracted algebra $\mathfrak{g}_0$ studied above.)
A \textit{resonant} expansion of $\mathfrak{g}$ with $S_E^{(N)}$ is then given by the space
\begin{align}
 \bigoplus_{i=0}^N  V_{i \,\text{mod}\,2} \otimes \lambda_{i}  
\end{align}
with the obvious Lie brackets. It is this refined version of a Lie algebra expansion that makes direct contact with~\eqref{eq:tan}. A simple example of the refined expansion would be to take the Poincar\'e algebra~\eqref{eq:Poin} and write it as
\begin{align}
\mathfrak{iso}(1,d) = \underbrace{\langle M_{ij}, P_0 \rangle}_{V_0} \oplus \underbrace{\langle M_{0i}, P_i\rangle}_{V_1}\,.
\end{align}
The expansion with $S_E^{(2)}$ then would have the basis elements
\begin{align}
M_{ij} \otimes \lambda_0\,, \quad P_0 \otimes \lambda_0 \,,\quad M_{0i}\otimes \lambda_1\,, \quad  P_i \otimes \lambda_1\,, \quad M_{ij}\otimes \lambda_2\,, \quad P_0 \otimes \lambda_2
\end{align}
with new non-trivial commutators
\begin{align}
[ M_{0i} \otimes \lambda_1 , M_{0j} \otimes \lambda_1 ] &= M_{ij} \otimes \lambda_2\,,\quad
[ M_{0i} \otimes \lambda_1 , P_j \otimes \lambda_1 ] = \delta_{ij} P_0 \otimes \lambda_2\,,\nn\\
[P_i\otimes \lambda_1 , P_j \otimes \lambda_1 ] &= 0
\end{align}
in the expanded algebra. 

The algebra obtained by expanding with $S^{(1)}_E$ gives the Galilei algebra~\eqref{eq:GalAlg}. The algebra above is a quotient of~\eqref{eq:infG}.
A more general discussion of the expansion method can be found in~\cite{Izaurieta:2006zz}.
We shall apply this method to several more cases in this paper.

\subsection{Free algebras, cohomology and quotients}
\label{eq:FLA}

We now turn to the discussion of free Lie algebras. General references for this are~\cite{Bourbaki:Lie,Viennot:1978} and we follow the exposition in~\cite{Gomis:2017cmt,Gomis:2019fdh}. 

A free Lie algebra on a (finite) set of $D=d+1$ generators 
\begin{align}
\label{eq:FLA1}
\mathfrak{f}_1 = \left\langle P_a \,\middle| \, a=0,1,\ldots d \right\rangle
\end{align}
is obtained by considering all possible multi-commutators of the generators $P_a$ only subject to anti-symmetry and the Jacobi identity. There is a natural grading of the free Lie algebra by the number of times the generators $P_a$ appear in the multi-commutator.\footnote{In later applications we shall also consider a refined double $\mathbb{Z}$-grading $(\ell,m)$ where the second label will be used to distinguish among the elements within $\mathfrak{f}_1$, see for instance section~\ref{sec:GFA}.} The infinite-dimensional free Lie algebra $\mathfrak{f}$ is therefore
\begin{align}
\label{eq:FLAdec}
\mathfrak{f} = \bigoplus_{\ell=1}^\infty \mathfrak{f}_\ell
\end{align}
with for example
\begin{align}
\mathfrak{f}_2 = \left\langle [P_a, P_b] \,\middle| \, a,b=0,1,\ldots,d \right\rangle = \ALT^2 \mathfrak{f}_1
\end{align}
being of dimension $\tfrac{D(D-1)}{2}$ because of the anti-symmetry of the commutator. We use the symbol $\ALT^k V$ to denote the $k$th anti-symmetric tensor power of a vector space $V$. The element $[P_a,P_b]$ is an independent element in the free Lie algebra.

The full structure of $\mathfrak{f}$ can be summarised elegantly by a generating series in a formal parameter $t$ as~\cite{Cederwall:2015oua,Gomis:2018xmo} 
\begin{align}
\bigotimes_{\ell=1}^\infty \left[ \bigoplus_{k=0}^\infty (-1)^k t^{k\ell}  \ALT^k \mathfrak{f}_\ell \right] = 1- t \mathfrak{f}_1
\end{align}
that leads for example to
\begin{align}
\label{eq:fex}
\mathfrak{f}_2 &= \ALT^2 \mathfrak{f}_1\,,\nn\\
\mathfrak{f}_3 &= \mathfrak{f}_1\otimes \mathfrak{f}_2 \ominus \ALT^3 \mathfrak{f}_1\,,\nn\\
\mathfrak{f}_3 &= \big( \mathfrak{f}_1\otimes \mathfrak{f}_3\oplus \ALT^2\mathfrak{f}_2\big) \ominus \mathfrak{f}_2 \otimes \ALT^2 \mathfrak{f}_1 \oplus \ALT^4 \mathfrak{f}_2\,.
\end{align}
With $\ominus$ we mean the removal of a vector space from the tensor product, so that for $\mathfrak{f}_3$ the formula states that one takes all commutators of $P_a$ from $\mathfrak{f}_1$ with the anti-symmetric $[P_b,P_c]$ from $\mathfrak{f}_2$ but has to remove the completely anti-symmetric Jacobi identity in all three elements. 

Free Lie algebras as defined above are graded consistently with~\eqref{eq:FLAdec},  i.e., they satisfy
\begin{align}
\label{eq:Fgrad}
[ \mathfrak{f}_\ell , \mathfrak{f}_m ] \subset \mathfrak{f}_{\ell+m}\,.
\end{align}

The elements in $\mathfrak{f}_\ell$ can be represented by Young diagrams with $\ell$ boxes that represent the irreducible action of the symmetric group $\mathcal{S}_D$ on the elements in a set of multicommutator. In this way we write
\begin{align}
\label{eq:Fyng}
\mathfrak{f}_1 \leftrightarrow \yng(1)\,,\quad\quad
\mathfrak{f}_2 \leftrightarrow \yng(1,1)\,,\quad\quad
\mathfrak{f}_3 \leftrightarrow \yng(2,1)\quad \text{etc.}
\end{align}

The free Lie algebra can also be viewed as a successive extension of the real commutative Lie algebra $\mathfrak{f}_1$ using the method of Chevalley--Eilenberg Lie algebra cohomology~\cite{Gomis:2017cmt}. The second cohomology of $\mathfrak{f}_1$ with values in $\mathbb{R}$ is non-trivial and of dimension $\tfrac{D(D-1)}2$ and therefore the Lie algebra $\mathfrak{f}_1$ can be extended by introducing anti-symmetric generators $Z_{ab}=Z_{[ab]}$ with the new commutator
\begin{align}
\label{eq:Max2}
[P_a, P_b] = Z_{ab}\,,
\end{align}
but the $Z_{ab}$ are central in this extended algebra. Thus one has obtained a graded Lie algebra $\mathfrak{f}_1\oplus\mathfrak{f}_2$ by considering the cohomology of $\mathfrak{f}_1$. The process can now be repeated by studying the cohomology of  $\mathfrak{f}_1\oplus\mathfrak{f}_2$ which leads to  $\mathfrak{f}_1\oplus\mathfrak{f}_2\oplus\mathfrak{f}_3$  and so on. In this way, the free Lie algebra $\mathfrak{f}$ is the maximal cohomological extension of $\mathfrak{f}_1$.

\medskip

As suggested by the notation~\eqref{eq:FLA1}, we wish to think of the elements of $\mathfrak{f}_1$ for instance as the translation generators of some kinematic algebra. Typically, there is also a set of rotation generators, such as the Lorentz generators $M_{ab}$, under which the translation generators form a module. We call the space of the rotation generators $\mathfrak{f}_0$ as then we have a graded structure
\begin{align}
\mathfrak{f}_0\oplus \mathfrak{f}_1
\end{align}
to begin with. As a Lie algebra this is a semi-direct sum since $\mathfrak{f}_0$ acts on its representation space $\mathfrak{f}_1$. The free Lie algebra based on $\mathfrak{f}_1$ then inherits an action of $\mathfrak{f}_0$ on each $\mathfrak{f}_\ell$ and the expressions in~\eqref{eq:fex} can be viewed as products and sums of $\mathfrak{f}_0$ modules. Extensions to super-algebras are discussed for example in~\cite{Cederwall:2015oua,Gomis:2018xmo}.

Free Lie algebras $\mathfrak{f}$ admit many different quotients. We list a few important and representative examples and consider the case when there are also rotations $\mathfrak{f}_0$ acting on the algebra, see also~\cite{Gomis:2017cmt}.
\begin{enumerate}
\item \underline{Level truncation:} Due to the grading~\eqref{eq:Fgrad}, the space 
\begin{align}
\mathfrak{i}_\ell = \bigoplus_{m>\ell} \mathfrak{f}_m
\end{align}
is a Lie algebra ideal inside $\mathfrak{f}$ for any integer $\ell>0$. The corresponding quotient 
\begin{align}
\mathfrak{q}_\ell = \mathfrak{f} / \mathfrak{i}_\ell\cong \bigoplus_{m=0}^\ell \mathfrak{f}_m
\end{align}
consists of all elements up to level $\ell$ (as a vector space) and commutators going beyond the truncation are set to zero.
\item \underline{Row truncation:}
Referring back to the representation~\eqref{eq:Fyng} of elements of $\mathfrak{f}$ as Young diagrams, we can define the space
\begin{align}
\mathfrak{s}_r = \langle \text{Young tableaux with more than $r$ rows} \rangle\,,
\end{align}
which is an ideal of $\mathfrak{f}$ since commutation only adds boxes but never removes them. The corresponding quotient $\mathfrak{r}_r= \mathfrak{f} / \mathfrak{s}_r$ then consists of all elements of $\mathfrak{f}$ with at most $r$ rows in their Young diagram.
\item \underline{Derivative truncation:}
The row truncation above can be refined by considering the ideal
\begin{align}
\mathfrak{u} = \left\langle \begin{array}{c}\text{Young tableaux with more than $2$ rows}\\\text{ or more than $2$ boxes in the second row}\end{array} \right\rangle\,.
\end{align}
The corresponding quotient 
\begin{align}
\mathfrak{d} = \mathfrak{f}/\mathfrak{u}
\end{align}
then consists only of those generators of $\mathfrak{f}$ whose Young diagrams have the shape
\begin{align}
\label{eq:derivq}
\yng(1)\,,\, \yng(1,1)\,,\, \yng(2,1)\,,\, \yng(3,1)\,,\, \yng(4,1)\,, \ldots
\end{align}
with an arbitrary number of boxes in the first row. Why we refer to this quotient as the derivative truncation will become clear in section~\ref{sec:Maxpart} below.
\end{enumerate}

Yet another common quotient is described by Serre relations and this arises for Kac--Moody algebras~\cite{Gabber:1981,Kac:1995} as we shall review in section~\ref{sec:KMA} below.

\subsubsection{Maxwell free Lie algebra}
\label{sec:Maxfree}

Let us illustrate the free Lie algebra construction in the simplest case where $\mathfrak{f}_0$ is the Lorentz algebra and $\mathfrak{f}_1$ are the translations $P_a$ of the Poincar\'e algebra~\eqref{eq:Poin}. In this case, we obtain the algebra
\begin{align}
[P_a , P_b ] = Z_{ab}\,,\quad\quad
[Z_{ab}, P_c ] = Y_{ab,c}\,,
\end{align}
where $Z_{ab}=Z_{[ab]}$ is a basis of $\mathfrak{f}_2$ and $Y_{ab,c}$ a basis of $\mathfrak{f}_3$. The symmetries of $Y_{ab,c}$ are those of the (2,1) hook:
\begin{align}
\label{eq:Young3}
Y_{ab,c} = Y_{[ab],c} \,,\quad\quad Y_{[ab,c]} = 0\,.
\end{align}
Under the Lorentz generators $M_{ab}$ all elements transform as tensors in the way that their indices dictate. 

The antisymmetric element $Z_{ab}$ arose first in studies of the extension of the Poincar\'e algebra in the presence of a constant electro-magnetic field~\cite{Bacry:1968zf,Bacry:1970ye,Schrader:1972zd} when one co-rotates the constant field $F_{ab}$ under Lorentz~\cite{Schrader:1972zd,Bonanos:2008ez}. The extension including the generators $Y_{ab,c}$ was also considered in~\cite{Bonanos:2008ez} where it was linked to linearly varying electro-magnetic backgrounds: $F_{ab} \sim Y_{ab,c} x^c$ (in Cartesian coordinates) and the Young irreducibility~\eqref{eq:Young3} is equivalent to the Bianchi identity $\partial_{[c} F_{ab]}\sim Y_{[ab,c]}=0$. The extension to the full free Lie algebra was considered in~\cite{Gomis:2017cmt}, see also section~\ref{sec:Maxpart} below for a corresponding particle model.

We note that one can also consider non-relativistic limits of the relativistic Maxwell algebra and there are different limits that arise depending on the scaling of the electric and magnetic fields. The corresponding algebras can be called electric, magnetic and pulse Maxwell algebras~\cite{Barducci:2019fjc,Gomis:2019fdh}.

\subsubsection{Galilean free Lie algebra}
\label{sec:GFA}

A second instance of the free Lie algebra construction can be obtained by starting from the Galilei algebra~\eqref{eq:GalAlg} and letting~\cite{Gomis:2019fdh}
\begin{align}
\label{eq:FGal}
\mathfrak{f}_0 = \langle J_{ij} \rangle\,,\quad 
\mathfrak{f}_1 = \langle B_i, H \rangle\,.
\end{align}
Note that this assignment of generators to levels is different from that that would be inherited directly from the Poincar\'e case in the previous section. However, this assignment is consistent with the grading due to the contracted commutation relations~\eqref{eq:GalAlg}. A consequence of~\eqref{eq:FGal} is that the translation generator $T_i$ occurs at level two in the free Lie algebra via the commutator 
\begin{align}
T_i = [H,B_i]\,.
\end{align}

\begin{table}[t]
\centering
\begin{tabular}{c|cccc}
& $\ell=0$ & $\ell=1$ &  $\ell=2$ &  $\ell=3$ \\\hline\\
$m=0$ & $J_{ij}$ & $B_i$ & $S_{ij}$ & $Y_{ij,k}$\\[2mm]
$m=1$ & & $H$ & $T_i$ & $B_{ij}$, $Z_{i,j}$ \\[2mm]
$m=2$ & &&& $Z_i$
\end{tabular}
\caption{\label{tab:FG}\sl The first few levels of the free Lie algebra generated by the (magnetic) Galilean choice~\eqref{eq:FGal}. The double-grading $(\ell,m)$ is explained in the text. A similar table has appeared in~\cite{Gomis:2019fdh}.}
\end{table}

The free Lie algebra generated from~\eqref{eq:FGal} was called the magnetic Galilei algebra in~\cite{Gomis:2019fdh} and it is the only case we consider here. Due to the presence of the rotation-invariant $H$ inside $\mathfrak{f}_1$, the structure of Young diagrams gets a bit more involved. The resulting structure actually admits a double-grading by assigning the level $(\ell,m)=(0,0)$ to the spatial rotations $J_{ij}$, the level $(1,0)$ to the Galilean boosts $B_i$ and level $(1,1)$ to the Hamiltonian $H$. The first few terms in the resulting free Lie algebra are then shown in table~\ref{tab:FG}.

The notation in the table is such that indices that are separated with commas are in separate columns of a Young diagram while unseparated ones are in the same column. For instance, the commutator between the Galilean boost $B_i$ and translation $T_i$ is
\begin{align}
[B_i, T_i ] = B_{ij} + Z_{i,j} 
\quad \text{with}\quad B_{ij}\leftrightarrow \yng(1,1)
\quad \text{and} \quad Z_{i,j} \leftrightarrow \yng(2)\,.
\end{align}
The symmetric tensor $Z_{i,j}$ can be traced using the Euclidean metric $\delta^{ij}$ and the corresponding scalar $M$ under rotations is nothing but the Bargmann central extension $[B_i,T_j]\propto \delta_{ij} M$. 

However, the free Lie algebra methods provides many further extensions of interest that are discussed in more detail in~\cite{Gomis:2019fdh}.

\subsubsection{Carrollian free Lie algebra}

In the same way as for the Galilei algebra above one can also construct a free Lie extension of the Carroll algebra~\eqref{eq:Calg}. By applying the duality $P_0\leftrightarrow P_i$ between the Carroll and Maxwell case discussed in section~\ref{sec:contr} one is led to starting from
\begin{align}
\mathfrak{f}_0 = \langle J_{ij} \rangle\,,\quad \mathfrak{f}_1 = \langle K_i, T_i \rangle
\end{align}
that should be compared to~\eqref{eq:FGal}.

\begin{table}[t]
\centering
\begin{tabular}{c|cccc}
& $\ell=0$ & $\ell=1$ &  $\ell=2$ &  $\ell=3$ \\\hline\\
$m=0$ & $J_{ij}$ & $K_i$ & $S_{ij}$ & $Y_{ij,k}$ \\[2mm]
$m=1$ & & $T_i$ & $B_{ij}$, $Z_{i,j}$ &$2{\times}\raisebox{-.2\height}{\scalebox{0.6}{\yng(2,1)}},\,\raisebox{-.2\height}{\scalebox{0.6}{\yng(1)}},\,\raisebox{-.2\height}{\scalebox{0.6}{\yng(3)}}$   \\[2mm]
$m=2$ & && $T_{ij}$ & $2{\times}\raisebox{-.2\height}{\scalebox{0.6}{\yng(2,1)}},\,\raisebox{-.2\height}{\scalebox{0.6}{\yng(1)}},\,\raisebox{-.2\height}{\scalebox{0.6}{\yng(3)}}$ \\[2mm]
$m=3$ & & & & $T_{ij,k}$
\end{tabular}
\caption{\label{tab:CG}\sl The first few levels of the free Lie algebra generated by Carroll. The double-grading $(\ell,m)$ is explained in the text, as is the relation between Young diagrams and comma-separated index notation.}
\end{table}

Running the free Lie algebra construction~\eqref{eq:fex} then produces as next generators the result shown in table~\ref{tab:CG}, where we also introduced a second grading $m$ to distinguish the generators $K_i$ and $T_i$. Some commutators defining the elements in the table are written explicitly as
\begin{align}
[ K_i, K_j ] &= S_{ij} \,,& [T_i, T_j] &= T_{ij}\,,&
[ T_i, K_j ] &= B_{ij} + Z_{i,j}\,,
\end{align}
where $Z_{i,j}$ is symmetric while all the other rank two tensors $S_{ij}$, $B_{ij}$ and $T_{ij}$ are anti-symmetric. The Carroll Hamiltonian $K$ (see~\eqref{eq:Calg}) is obtained as the trace of the symmetric tensor:
\begin{align}
\delta^{ij} [ T_i, K_j] =  \delta^{ij} Z_{i,j}  =  d K\,,
\end{align}
where we recall that $d$ is the number of spatial dimensions.

We can recover the infinite Carroll algebra~\eqref{eq:infC} from the free Lie algebra construction by following steps similar to~\cite{Gomis:2019fdh}. By restricting to  anti-diagonal lines (of fixed $\ell-m$) in table~\ref{tab:CG}, restricting further to $m\in \{0,1\}$ and keeping only generators of vector and scalar type under rotations, we obtain an infinity of generators $J_{ij}^{(n)}$, $K_i^{(n)}$, $T_i^{(n)}$ and $K^{(n)}$ whose commutation relations are those of~\eqref{eq:infC}. Besides $K^{(0)} \sim \delta^{ij} Z_{i,j}$ we also have for example $J_{ij}^{(1)} \sim S_{ij}$ and $K_i^{(1)} \sim \delta^{jk} Y_{ij,k}$. As the infinite Carroll algebra~\eqref{eq:infC} is an expansion of the the Poincar\'e algebra, we see that the Carrollian free Lie algebra contains this particular expansion as a quotient.

\subsection{Connection to Kac--Moody algebras}
\label{sec:KMA}

In this final section on algebraic construction we would like to make a brief comment on the relation to (affine) Kac--Moody algebras. For any finite-dimensional Lie algebra $\mathfrak{g}$ it is well-known that one can construct the (untwisted) loop algebra by letting
\begin{align}
\hat{\mathfrak{g}} = \mathfrak{g}[\lambda,\lambda^{-1}]
\end{align}
of Laurent polynomials in $\lambda$ with values in $\mathfrak{g}$. It is also possible to add a central term and a derivation element to this construction to obtain a proper Kac--Moody algebra~\cite{Kac:1995}. 

The relation to the constructions above becomes transparent by restricting to the parabolic subalgebra of polynomials  $\mathfrak{g}[\lambda]$ whose elements can be written in terms of the basis $t_\alpha \otimes \lambda^n$ for $n=0,1,\ldots$. Clearly, this can be seen as a version of the method~\eqref{eq:tan} when setting all offsets $n_0(\alpha)$ to zero. Setting some of the offsets to a non-zero value can result in twisted in affine algebras, see~\cite{Gomis:2019fdh} for examples.

The parabolic subalgebra $\mathfrak{g}[\lambda]$ is also closely related to free Lie algebras. Indeed it is known that the Borel subalgebra of a Kac--Moody algebra can be described as the quotient of a free Lie algebra on the simple Chevalley generators (often denoted $e_i$) subject to the Serre relations that are encoded in the generalised Cartan matrix of the Kac--Moody algebra~\cite{Gabber:1981}. This is in particular true for affine algebras. Since we are dealing with a parabolic subalgebra rather than a Borel subalgebra in that the starting finite-dimensional Lie algebra $\mathfrak{g}$ is not the abelian Cartan subalgebra, the Serre relations have to be adapted slightly but one can still describe $\mathfrak{g}[\lambda]$ as a quotient of the free Lie algebra with $\mathfrak{f}_1=\mathfrak{g}$ that is acted upon by $\mathfrak{f}_0=\mathfrak{g}$. Again, refinements of this constructions are available when $\mathfrak{g}$ is decomposed already into $V_0\oplus V_1$, see~\cite{Gomis:2019fdh} for examples.\footnote{In this case a relation of the Galilean construction to the $\mathfrak{B}_N$ algebras introduced in~\cite{Salgado:2013eut} has been noticed in~\cite{Gomis:2019fdh}.} This shows that all the various algebraic constructions in this section are interrelated.


\section{Geometric realisations}
\label{sec:geo}

Suppose we have a kinematic algebra $\mathfrak{g}$ that has an action on some space-time $M$. For a given contraction $\mathfrak{g}_0$ of $\mathfrak{g}$ with corresponding perturbative expansion $\mathfrak{g}^{(\infty)}$ as introduced in the section~\ref{sec:contr}, we now want to construct a space $M^{(\infty)}$ on which this algebra acts.\footnote{Our considerations will be purely local and leave out questions of topology of the spaces.} Since $\mathfrak{g}^{(\infty)}$ is interpreted as the perturbative expansion of the kinematic algebra in the parameter $\lambda$, this will provide the arena to describe perturbatively corrected dynamics.

Let $x^a$ denote a set of local coordinates of $M$ on which there is a faithful action of $\mathfrak{g}$. This set of coordinates is assumed to be compatible with the contraction to $\mathfrak{g}_0$ in the sense that one can define an associated contraction on the coordinates such that $\mathfrak{g}_0$ acts faithfully on the contracted coordinates. 

With this we mean that
\begin{align}
x^a_{(n)} = x^a \otimes \lambda^{-n(a) - n}
\end{align}
with an appropriate offset $n(a)$ depending on the coordinate and restricted to $n=0$ carries a faithful action of $\mathfrak{g}_0=\mathfrak{g}^{(0)}$. The space $M^{(\infty)}$ on which $\mathfrak{g}^{(\infty)}$ acts is then given by considering all values of $n\geq 0$. 
The opposite sign of $n$ on the coordinates compared the generators in~\eqref{eq:tan} is due to the fact that we would like formal Lie algebra elements
\begin{align}
\sum_{n\geq 0} x^a_{(n)} P_a^{(n)} 
\end{align}
to be dimensionless which means that the scaling of $x^a_{(n)}$ with $\lambda$ must be opposite to that of $P_a^{(n)}$ (whose precise definition depends on the context). 
The Lie algebra element above can then be exponentiated and used in a non-linear realisation.
We now exemplify these considerations in various cases. We shall also construct a case of a generalised space-time related to a full free Lie algebra rather than to an expanded algebra.

\subsection{Post-Galilean space-time}

For the Poincar\'e algebra~\eqref{eq:Poin} the space $M$ is Minkowski space in $D$ dimensions with coordinates $x^a$ of dimension $L=\text{length}$. 
The faithful action of the Poincar\'e algebra can described as follows. Let 
\begin{align}
\frac12 \omega^{ab} M_{ab} + \alpha^a P_a
\end{align}
be an arbitrary element of the Poincar\'e algebra. Its action on the coordinate $x^a$ is given by
\begin{align}
\label{eq:PoinX}
\delta x^a = -\omega^{ab} \eta_{bc} x^c + \alpha^a\,.
\end{align}
For the Lorentz part $\mathfrak{so}(1,d)$ this is nothing but the fundamental representation on which the $P_a$ act as translations.

For the Galilean contraction~\eqref{eq:Gcontrlevy} we split space and time $a=(0,i)$ and let
\begin{align}
\label{eq:MGinf}
t_{(n)} = x^0 \otimes \lambda^{-n} \,, \quad x^i_{(n)} = x^i \otimes \lambda^{-n-1/2}\,.
\end{align}
Here, the dimensions of $t_{(n)}$ and $x^i_{(n)}$ are fixed by the dimension of $x^0$ and $x^i$ (that we always maintain at dimension length) and that of $\lambda$ which for the Galilean case follows from $\lambda=c^{-2}$.
In particular, even though we use the notation $t_{(m)}$, the lowest element $t_{(0)}$ does not have the dimension of time ($T$) but of length ($L$) and $x_{(0)}^i$ does not have the dimension of length but of $L^2/T$. 
We note also that in our conventions the Poincar\'e generators $P_a$ in~\eqref{eq:Poin} have dimension of ${L}^{-1}$ while the $M_{ab}$ are dimensionless.

The action of an element 
\begin{align}
\sum_{m\geq 0}\left[  \frac12 \omega^{ij}_{(m)} J_{ij}^{(m)}  + v^i_{(m)} B_i^{(m)} + \alpha^i_{(m)} T_i^{(m)} + \epsilon_{(m)} H^{(m)} \right]
\end{align}
of the algebra~\eqref{eq:infG} on a (dimensionless) coordinate element
\begin{align}
\sum_{n\geq 0} \Big[ x^i_{(n)} T_i^{(n)} + t_{(n)} H^{(n)} \Big]
\end{align}
 is then given by the commutator of the two elements, leading to
\begin{align}
\delta t_{(n)} &=  \epsilon_{(n)} + \sum_{m=0}^{n-1} \delta_{ij} v^i_{(m)} x^j_{(n-1-m)}\,,\\
\delta x^i_{(n)} &=  \alpha^i_{(n)} + \sum_{m=0}^{n}  \Big( -\omega^{ij}_{(m)} \delta_{jk} x^k_{(n-m)} + v^i_{(m)} t_{(n-m)}\Big)\,.
\end{align}
We see that restricting to only level $0$ this becomes the usual action on the Galilean coordinates $(t,z^i)$ with $t=t_{(0)}$ and $z^i = x^i_{(0)}$. In particular, the Galilean boost with only $v^i_{(0)}\neq 0$ yields
\begin{align}
\delta t = 0 \,,\quad \delta z^i = v^i_{(0)} t\,,
\end{align}
which is the lowest order term of the Lorentz boost. Note that in our conventions the parameter $v_{(0)}$ has dimension of $L/T$ as a velocity, but we recall that $[x^i_{(0)}]=L^2/T$ and $[t_{(0)}]=L$.

In order to see the systematic higher order expansion of the Lorentz boost encoded in~\eqref{eq:MGinf}, we follow~\cite{Gomis:2019sqv} and define collective coordinates formally by
\begin{align}
\label{eq:Xcol}
X^0 = \sum_{n\geq 0} t_{(n)} \lambda^{n} \,,\quad X^i = \sum_{n\geq 0} x^i_{(n)} \lambda^{n+1/2}\,,
\end{align}
as well as the collective boost parameter
\begin{align}
\Theta^i = \sum_{n\geq 0} v^i_{(n)}  \lambda^{n+1/2}\,.
\end{align}
The transformation of the collective coordinates~\eqref{eq:Xcol} under such a collective boost then works out as
\begin{align}
\label{eq:boost1}
\delta X^0 = \delta_{ij}  \Theta^i X^j \,,\quad
\delta X^i = \Theta^i X^0\,,
\end{align}
the usual expression for an infinitesimal relativistic Lorentz boost with rapidity $\Theta^i$. However, the difference is that now the boost parameter and the coordinate are collective. 

If one imposes that
\begin{align}
\label{eq:boostp1}
v^i_{(n)} = \frac{1}{2n+1} v^{2n+1} n^i
\end{align}
for some scalar $v$ and spatial unit vector $n^i$, i.e., $\delta_{ij} n^i n^j=1$, then the transformations~\eqref{eq:boost1} become for $\lambda=c^{-2}$
\begin{align}
\label{eq:boostG}
\delta X^0 = \sum_{n\geq 0} \frac{1}{2n+1} \left(\frac{v}{c}\right)^{2n+1} \delta_{ij} n^i X^j\,,\quad
\delta X^i = \sum_{n\geq 0} \frac{1}{2n+1} \left(\frac{v}{c}\right)^{2n+1} n^i X^0\,,
\end{align}
which are the expansions of the infinitesimal Lorentz boost with parameter $\theta^i = \theta n^i$, where $\tanh \theta=v/c$ for $v/c\ll 1$.

At this point we should comment on the geometrical meaning of the collective coordinates~\eqref{eq:Xcol}. These define a hyperspace of co-dimension $D$ within the infinite-dimensional generalised Minkowski space with coordinates~\eqref{eq:MGinf}. Since the sums are infinite and we are not making any assumptions about convergence here, the expressions are formal but the formal expansion parameter $\lambda$ is introduced in such a way as to render meaningful expressions at any finite order in the expansion. 
What the transformation~\eqref{eq:boost1} then describes is a transformation from one hyperspace to another one, so we obtain a description of ordinary Minkowski space as a family of hyperspaces inside generalised Minkowski space. We shall see that a similar picture applies to all other expansions considered in this paper.

\subsection{Post-Carrollian space-time}
\label{sec:CarST}

For the case of the Carroll algebra, we use `Carroll time' 
$s=  C x^0$ introduced in~\cite{LevyLeblond:1965,Duval:2014uoa}. The contraction limit in these variables is $C\to\infty$. Morally, we can think of $C$ as being related to the inverse of the speed of light, so that the speed of light goes to zero. However, the dimension of $C$ is that of a velocity. The expansion parameter $\lambda=C^{-2}$, so that $s_{(0)} = x^0 \otimes C$ is the Carroll time of~\cite{LevyLeblond:1965,Duval:2014uoa}.

For the Carrollian contraction~\eqref{eq:Ccontr} we proceed analogously to the generalised Galilei space-time and define
\begin{align}
\label{eq:MCinf}
s_{(n)} = x^0 \otimes \lambda^{-n-1/2} \,,\quad x^i_{(n)} = x^i \otimes \lambda^{-n}\,.
\end{align}
where the difference to~\eqref{eq:MGinf} that the constant shift has moved from the space to the time translations.  The dimensions of the coordinates implied by these definitions are $[s_{(n)}] = L^{2n+2}/T^{2n+1}$ and $[x_{(n)}^i] = L^{2n+1}/T^{2n}$.

A dimensionless element 
\begin{align}
\sum_{m\geq 0} \Big[ \frac12 \omega^{ij}_{(m)} J_{ij}^{(m)}  + v^i_{(m)} K_i^{(m)} + \alpha^i_{(m)} T_i^{(m)} + \epsilon_{(m)} K^{(m)} \Big]
\end{align}
of the expanded Carroll algebra~\eqref{eq:infC} then acts on the coordinates by 
\begin{align}
\delta s_{(n)} &=  \epsilon_{(n)} + \sum_{m=0}^{n} \delta_{ij} v^i_{(m)} x^j_{(n-m)}\,,\\
\delta x^i_{(n)} &=  \alpha^i_{(n)} - \sum_{m=0}^{n}  \omega^{ij}_{(m)} \delta_{jk} x^k_{(n-m)} + \sum_{m=0}^{n-1} v^i_{(m)} s_{(n-1-m)}\,.
\end{align}

Especially, restricting to lowest order we obtain for the Carrollian time coordinate $s=s_{(0)}$ and $z^i=x^i_{(0)}$ that the Carrollian boost (only $v^i_{(0)}\neq 0$) acts by
\begin{align}
\delta s = \delta_{ij} v^i_{(0)}  z^j\,,\quad
\delta z^i = 0\,,
\end{align}
the well-known expression for this boost, see e.g.~\cite{LevyLeblond:1965,Duval:2014uoa}.
In particular, an ordinary particle at rest cannot be Carroll boosted to one in motion: it is effectively stationary in any frame.

We now turn to corrections to this classical statement as contained in the infinite-dimensional algebra~\eqref{eq:infC}.
We introduce the collective coordinates
\begin{align}\label{carrollexpansion}
X^0 = \sum_{n\geq 0} s_{(n)} \lambda^{n+1/2} \,,\quad
X^i = \sum_{n\geq 0} x^i_{(n)} \lambda^{n}
\end{align}
as well as the collective boost parameter 
\begin{align}
\Theta^i = \sum_{n\geq 0} v^i_{(n)} \lambda^{n+1/2}\,.
\end{align}
The transformation then becomes
\begin{align}
\label{eq:boost2}
\delta X^0 =  \delta_{ij} \Theta^i X^j \,,\quad
\delta X^i = \Theta^i X^0
\end{align}
just as in~\eqref{eq:boost1} and thus formally resembles the usual infinitesimal Lorentz boost with parameter $\Theta^i$. With $\lambda=C^{-2}$ we can now specialise to
\begin{align}
v^i_{(n)} = \frac1{2n+1} b^{2n+1} n^i
\end{align}
to arrive at
\begin{align}
\delta X^0 = \sum_{n\geq 0} \frac{1}{2n+1} \left(\frac{b}{C}\right)^{2n+1} \delta_{ij} n^i X^j\,,\quad
\delta X^i = \sum_{n\geq 0} \frac{1}{2n+1} \left(\frac{b}{C}\right)^{2n+1} n^i X^0\,.
\end{align}
This is the correct expansion of a Lorentz boost in Carroll parametrisation where $b= C \frac{v}{c}$ is fixed in the limit $C\to \infty$~\cite{Duval:2014uoa}.

\subsection{Conformal post-Galilean space-time}
\label{sec:GCA}

The relativistic conformal algebra in $D>2$ dimensions is $\mathfrak{so}(D,2)$ that contains, besides the Poincar\'e generators~\eqref{eq:Poin}, the special conformal generators $S_a$ and the dilatation generator $D$ with additional commutators 
\begin{align}
[ M_{ab} , S_c ] &= \eta_{bc} S_a - \eta_{ac} S_b\,,\quad [D , P_a] = P_a \,,\quad [D, S_a] = - S_a\,,\nn\\
[ S_a, P_b ] &=  2 M_{ab} - 2\eta_{ab} D\,.
\end{align}
A non-relativistic Galilean version of this algebra can be obtained by considering the contraction ($\lambda=c^{-2}$)
\begin{align}
\label{eq:contrGCA}
J_{ij} &= M_{ij}\,, & \mathcal{D} &= D \,,& H&= P_0\,,& S&=S_0\,,\nn\\
B_i &= \lambda^{1/2} M_{i0} \,,& T_i &=\lambda^{1/2} P_i\,,&  G_i &=\lambda^{1/2} S_i
\end{align}
that extends the Galilean contraction~\eqref{eq:Gcontrlevy} from Poincar\'e to the conformal algebra. The resulting contracted algebra is known as the Galilei conformal algebra and has been studied for example in~\cite{negro1997nonrelativistic,Henkel:1997zz,Lukierski:2005xy,Bagchi:2009my,Duval:2011mi}, see also~\cite{Ammon:2020fxs,Campoleoni:2021blr} for a recent extension to higher spin algebras.

The two lines of~\eqref{eq:contrGCA} also define spaces $V_0$ and $V_1$ satisfying~\eqref{eq:expC}, so that an infinite expanded algebra undoing the contraction can be defined, exactly in the same way as for the previous cases.  An infinite generalised space-time on which the infinite expanded Galilei conformal algebra can act is then defined by introducing coordinates
\begin{align}
t_{(n)} = x^0 \otimes \lambda^{-n} \,,\quad x_{(n)}^i = x^i \otimes \lambda^{-n-1/2}\,.
\end{align}
The action of the algebra on these coordinates can be worked out in the same way as in the previous cases, with the additional feature that the action of the special conformal transformation is non-linear in the coordinates
due to the relativistic expressions
\begin{align}
\label{eq:confX}
\delta_{\sigma D}  x^a &= \sigma x^a\,,& \delta_{\beta^b S_b} x^a &= 2 (x\cdot \beta) x^a - (x\cdot x) \beta^a\,,
\end{align}
extending the Poincar\'e transformations~\eqref{eq:PoinX}.

Collectives coordinates are defined by
\begin{align}
\label{eq:Xconf}
X^i = \sum_{n\geq 0} x_{(n)}^i \lambda^{n+1/2}\,, \quad X^0 = \sum_{n\geq 0} t_{(n)} \lambda^n
\end{align}
exactly as for the non-conformal Galilei case~\eqref{eq:Xcol}. Under special conformal transformations~\eqref{eq:confX}, the lowest order coordinates transform as
\begin{align}
\label{eq:expconf}
\delta t_{(0)} &= -b_{(0)} t_{(0)}^2 \,,\nn\\
\delta t_{(1)} &= -2b_{(0)} t_{(0)} t_{(1)} - b_{(1)} t_{(0)}^2 +2 \delta_{ij} \beta_{(0)}^i x_{(0)}^j t_{(0)} - \delta_{ij} x_{(0)}^i x_{(0)}^j b_{(0)}\,,\nn\\
\delta x_{(0)}^i &= -2 b_{(0)} t_{(0)} x_{(0)}^i + t_{(0)}^2 \beta_{(0)}^i\,,
\end{align}
where we have expanded the parameter of the transformation as
\begin{align}
\beta^0 = \sum_{n\geq 0} b_{(n)} \lambda^n \,,\quad \beta^i =\sum_{n\geq 0} \beta^i_{(n)} \lambda^{n+1/2}\,.
\end{align}

\subsection{Post-Minkowski space-time}

The small parameter can also be taken to be the curvature of space-time in appropriate dimensions. This was considered in~\cite{Gomis:2020wrv} and leads to corrections to Minkowski space-time towards (Anti-)de Sitter space when the starting point is the (A)dS algebra that differs from the Poincar\'e algebra~\eqref{eq:Poin} by the non-trivial commutator 
\begin{align}
\label{eq:AdSalg}
[ P_a, P_b] = \sigma M_{ab}
\end{align}
among the translations. The sign $\sigma=+1$ is the AdS algebra $\mathfrak{so}(D-1,2)$ and $\sigma=-1$ is the dS algebra $\mathfrak{so}(D,1)$.\footnote{The AdS algebra in $D$ dimensions is famously isomorphic to the conformal algebra in $D-1$ dimensions. Since we use the indices $a$ to run over the space-time dimension, the range of indices in this section and section~\ref{sec:GCA} is different although they are based on the same types of algebra. However, they also address different contractions and expansions.} Here, and in contrast to the Poincar\'e algebra~\eqref{eq:Poin}, we have rescaled all generators to be dimensionless.\footnote{If one wanted to keep the dimensions of $P_a$ at $L^{-1}$ this would require keeping an explicit $1/R^2$ on the right-hand side of the commutator~\eqref{eq:AdSalg}, where $R$ is the (A)dS radius.}

Following our usual expansion method we define the generators
\begin{align}
\label{eq:AdSexp}
M_{ab}^{(n)} = M_{ab} \otimes \lambda^n \,, \quad P_a^{(n)} = P_a \otimes \lambda^{1/2+n}\,,
\end{align}
where now $\lambda= R^{-2}$ is to be thought of as the curvature scale of the (A)dS space-time. For $R\to\infty$ we obtain the Poincar\'e algebra~\eqref{eq:Poin} as a contraction of the (A)dS algebra similar to the non-relativistic cases in section~\ref{sec:contr}.

One can now similarly consider an extended space-time with coordinates
\begin{align}
\label{eq:AdSco}
x_{(n)}^a = x^a \otimes \lambda^{-1/2-n}\,.
\end{align}
The transformations formula for these coordinates is now more complicated since the underlying translations no longer commute due to~\eqref{eq:AdSalg}. Since we do not rely on them in the following, we refer the reader to~\cite{Gomis:2020wrv}. In section~\ref{sec:AdSpart} we shall study a particle model based on this generalised space-time.

\subsection{Minkowski--Maxwell space-time}

In the case of the Maxwell extension of Poincar\'e we also deal with non-commuting translations $P_a$, the basic commutator is~\eqref{eq:Max2}, where $Z_{ab}$ is a new generator unlike in the case of the (A)dS algebra.

The most general algebra that we can construct when starting from the Poincar\'e algebra is the Maxwell free Lie algebra that was introduced in section~\ref{sec:Maxfree}.  An associated generalised space-time can be defined by considering the $P_a$ and all their free commutators as translation generators. This means that one has coordinates for each of them~\cite{Bonanos:2008ez,Gomis:2017cmt}
\begin{align}
\label{eq:Maxco}
P_a \leftrightarrow x^a \,,\quad  Z_{ab} \leftrightarrow \theta^{ab}\,,\quad Y_{ab,c} \leftrightarrow \xi^{ab,c} \quad \text{etc.}
\end{align}
The generalised space-time defined by these coordinates has non-abelian translations
\begin{align}
\delta x^a &= \epsilon^a\,,\nn\\
\delta \theta^a &= \epsilon^{ab} - \frac12 (x^a \epsilon^b - x^b \epsilon^a)\,,\\
\delta \xi^{ab,c} &= \epsilon^{ab,c} +\frac13\left( 2\epsilon^{ab} x^c- \epsilon^{bc} x^a - \epsilon^{ca} x^b\right) +\frac13\left( \epsilon^a x^b x^c -\epsilon^b x^a x^c\right)\,,\nn
\end{align}
where the higher-level coordinates are also affected by the translations of all lower levels.

In section~\ref{sec:Maxpart}, we consider a particle model on the associated space-time and how it relates to the motion of charged particle in an electro-magnetic background field. We also note that one can consider various non-relativistic limits of Maxwell algebras and space-times~\cite{LeBellac:1973,Barducci:2019fjc,Gomis:2019fdh}.

\section{Free actions}
\label{sec:act}

In this section, we consider particle actions for free spinless particles in the various generalised space-times constructed in the previous section. We shall discuss in particular how they can be used to reproduce the corrections to the usual relativistic free particles. The case of the Carrollian generalisation will be discussed in most detail since it is less well-covered in the literature but has recently attracted attention in the context of cosmology and gravity~\cite{Bergshoeff:2017btm,Gomis:2019nih,Henneaux:2021yzg,deBoer:2021jej,Hansen:2021fxi}. We shall consider both tachyonic and ordinary particles and the resulting corrections in the case of Carroll are new to the best of our knowledge. In general, we shall parametrise the world-lines of particles using a parameter $\tau$ and denote derivatives with respect to this parameter by dots. The dimension of this parameter will be that of time ($T$) for Galilei but that of Carroll time ($L^2/T$) for Carroll.

\subsection{Particle in post-Galilean space-time}
\label{sec:Gpart}

The starting point for all actions comes from the expansion of the relativistic invariant metric using the collective coordinates~\eqref{eq:Xcol}
\begin{align}
ds^2 = \eta_{ab} dX^a dX^b = \sum_{m,n\geq 0} \lambda^{m+n} \left( -  dt_{(m)} dt_{(n)} + \lambda \delta_{ij} \,dx^i_{(m)} dx^j_{(n)} \right)\,,
\end{align}
where the factor of $\lambda$ in front of the spatial metric is crucial.

We shall first consider the usual massive relativistic particle, corresponding to a time-like norm of the velocity vector, and its Galilean limit. Then we shall consider the same procedure for a relativistic tachyon whose velocity vector is space-like and whose Galilean limit is a massless Galilean particle. The intuitive reason for this is that massless propagation in Newtonian physics is instantaneous which corresponds to space-like trajectories in Minkowski space. Galilean limits of relativistic light-like particles will not be considered in the context of post-Galilean space-time but in its conformal extension in section~\ref{sec:GCApart}.

\subsubsection{Massive Galilean particle}
\label{sec:MGP}

Perturbative actions for a massive particle can be obtained by expanding the reparametrisation invariant configuration space action
\begin{align}
\label{eq:PAGal}
S = - m \lambda^{-1/2} \int d\tau \sqrt{-\eta_{ab} \dot{X}^a \dot{X}^b} = S_{(0)} + S_{(1)} + S_{(2)} + \ldots
\end{align}
in powers of $\lambda=c^{-2}$ with the result 
\begin{align}
S_{(0)} &= -m\lambda^{-1/2} \int d\tau \, \dot{t}_{(0)}\,,\nn\\
S_{(1)} &= m \lambda^{1/2}\int d\tau \Big( -\dot{t}_{(1)} + \frac{\delta_{ij} \dot{x}^i_{(0)} \dot{x}^j_{(0)}}{2\dot{t}_{(0)}}\Big)\,,\nn\\
\label{eq:S2G}
S_{(2)} &= m \lambda^{3/2} \int d\tau \Big( -\dot{t}_{(2)} +\frac{\delta_{ij} \dot{x}^i_{(0)} \dot{x}^j_{(1)}}{\dot{t}_{(0)}} - \dot{t}_{(1)} \frac{ \delta_{ij} \dot{x}^i_{(0)} \dot{x}^j_{(0)}}{2\dot{t}_{(0)}^2} + \frac{(\delta_{ij}\dot{x}^i_{(0)}\dot{x}^j_{(0)})^2}{8 \dot{t}_{(0)}^3}\Big)
\end{align}
This was given in~\cite{Gomis:2019sqv} up to the fact that the dimensions of the variables here differ from there by a factor of $c$. 

The actions written in~\eqref{eq:S2G} have global symmetries associated with the expanded algebras from section~\ref{sec:contr} up to the order in the expansion. In particular, the action $S_{(2)}$ has more symmetries than the usual Galilei (or Bargmann) invariance. In addition, the actions have gauge symmetries generated by first-class constraints~\cite{Gomis:2019sqv}. Gauge-fixing these symmetries one still retains enhanced global symmetries that are realised non-linearly. As we shall describe next, we will also identify the space coordinates $x_{(m)}^i$ to a single $x^i$ through what we refer to as `choosing a slice'. This step makes it possible to connect to the usual non-relativistic expansions at the price of breaking the global symmetries. 

\medskip

The first action $S_{(0)}$ is a total derivative and does not describe any non-trivial local dynamics. The existence of this term is nevertheless significant and related to the possibility of centrally extending the Galilei algebra to the Bargmann algebra~\cite{levy1969group}.

The next action $S_{(1)}$  becomes the usual non-relativistic $\frac12 m \dot{x}^2$ after gauge-fixing $t_{(0)} = c\tau$ and dropping a total derivative. 
Moreover, we identify as a slice condition
\begin{align}
x^i_{(0)} = c x^i
\end{align}
in order to obtain conventional dimensions, but we emphasise that this step is \textit{not} fixing a gauge symmetry but breaks symmetries~\cite{Gomis:2019sqv}.

For the next order term $S_{(2)}$, we similarly gauge-fix 
\begin{equation}
\label{eq:GalGF}
t_{(0)}= c\tau\quad \text{and} \quad t_{(1)}=c^3 \tau\,
\end{equation}
and choose a slice as
\begin{equation}
\label{eq:GalSC}
x_{(0)}^i = c x^i \quad \text{and} \quad x_{(1)}^i  = c^3 x^i\,.
\end{equation}
This choice of slice is dictated by the relation~\eqref{eq:MGinf} between Minkowski and generalised Minkowski space.
Plugging this into~\eqref{eq:S2G} leads to (using vector notation for the spatial components for simplicity)
\begin{align}
\label{eq:S2Gnormal}
\tilde{S}_{(2)} =   \int d\tau \left(-mc^2 +  \frac12 m\dot{\vec{x}}^{\, 2} + \frac{m}{8c^2} (\dot{\vec{x}}^{\, 2})^2\right)\,.
\end{align}
Working out the energy of the particle associated with this action leads to
\begin{align}
E = mc^2 + \frac12 m \dot{\vec{x}}^{\, 2} + \frac{3m}{8c^2} (\dot{\vec{x}}^{\, 2})^2
\end{align}
which agrees with the expansion of the relativistic energy
\begin{align}
E = \frac{mc^2}{ \sqrt{1 - \frac{\dot{\vec{x}}^{\, 2}}{c^2}}}
\end{align}
to the order given. Such an analysis can be performed to any desired order. Similarly, the momentum can be worked out as
\begin{align}
    \vec{P}= m \dot{\vec{x}} + \frac{m}{2c^2} (\dot{\vec{x}}^{\, 2}) \dot{\vec{x}} \,,
\end{align}
which are the first two terms in the large-$c$ expansion of $ \vec{P} = m \dot{\vec{x}} / \sqrt{1-\dot{\vec{x}}^{\,2}/c^2}$.
Note that in order to get the desired expansion to order $n-1$ we only have to consider the action $S_{(n)}$ that comprises all corrections to that order. 
The action~\eqref{eq:S2Gnormal} has no global symmetries but our procedure allowed us to systematically arrive at it through a formalism with enhanced symmetry.

There is an ambiguity in interpreting the actions~\eqref{eq:S2G} in terms of which space they are defined on. Since we started with an action on Minkowski space it is natural to view the actions as being defined on generalised Minkowski space to the same order in $\lambda$ for both time and space variables. This means for example that we would like to view 
\begin{align}
S_{(0)} = S_{(0)}(t_{(0)}, x_{(0)}) =  - m \lambda^{-1/2} \int d\tau \, \dot{t}_{(0)}
\end{align}
to depend also formally on $x_{(0)}$ although $x_{(0)}$ does not enter the action at all. Taking this point of view implies that there are \textit{two} canonical constraints associated to the action $S_{(0)}$, namely
\begin{align}
E_{(0)} = - \frac{\partial L_{(0)}}{\partial \dot{t}_{(0)}} = m \lambda^{-1/2}\,,\quad\quad
\vec{p}_{(0)} = \frac{\partial L_{(0)}}{\partial \dot{\vec{x}}_{(0)}}   =0\,.
\end{align}
The second constraint $\vec{p}_{(0)}=0$ can be viewed as somewhat artificial mathematically but stems from the physical origin of Minkowski space. In a similar way, the $N$th order action $S_{(N)}$ will always contain an extra constraint $\vec{p}_{(N)}=0$ by making $S_{(N)}$ depend on the same number $N+1$ of $t_{(n)}$ and $\vec{p}_{(n)}$.

\subsubsection{Massless Galilean particle}
\label{sec:M0G}

The massless Galilean particle~\cite{Souriau} can be obtained as the non-relativistic limit of the relativistic tachyon~\cite{Batlle:2017cfa}. From the point of view of the kinematic algebra the massless Galilean particle has vanishing Bargmann central charge.
We will work out the first correction starting from a phase form of the action, starting from 
\begin{align}
\label{eq:m0G}
    S= \int d\tau \left[ -E \dot{T} + \vec{P} \cdot \dot{\vec{X}} - \frac{e}{2} \left( -\frac{E^2}{c^2} + \vec{P}^{\, 2} -k^2 \right)\right]\,,
\end{align}
where we have introduced the `colour' $k^2=m^2c^2$~\cite{Souriau}. The lowest order term in the limit $c\to \infty$ removes the energy from the mass-shell constraint, leading to the action~\cite{Batlle:2017cfa}
\begin{align}
\label{eq:massG0}
     S_{(0)}= \int d\tau \left[ -E_{(0)} \dot{t}_{(0)} + \vec{p}_{(0)} \cdot \dot{\vec{x}}_{(0)} - \frac{e_{(0)}}{2} \left(  \vec{p}_{(0)}^{\, 2} -k^2 \right)\right]\,.
\end{align}
Here, we have used the expansions ($\lambda = c^{-2}$) 
\begin{align}
\label{eq:expPS}
   T &= \sum_{n\geq 0} t_{(n)} \lambda^{n}\,,&
    X^i &= \sum_{n\geq 0} x^i_{(n)} \lambda^{n}\,,\nn\\
    E &= \sum_{n\geq 0} E_{(n)} \lambda^{n}\,,&
    P^i &= \sum_{n\geq 0} p^i_{(n)} \lambda^{n}\,,&
    e &= \sum_{n\geq 0} e_{(n)} \lambda^n\,.
\end{align}
These expansions are consistent with~\eqref{eq:Xcol} except for an adjustment of dimensions. The variables $(t_{(0)},\, x^i_{(0)},\, E_{(0)},\, p^i_{(0)})$ appearing here have canonical dimensions while the ones in~\eqref{eq:Xcol} were rescaled by factors of $c$. If one wanted to use the expansion~\eqref{eq:Xcol} directly while keeping the collective symplectic term $\dot{X}^i P_i$ free of factors of $\lambda$, this would also require rescaling $P_i$ and therefore introduce additional factors of $\lambda$ into the collective mass-shell constraint.

One thing we can immediately deduce from~\eqref{eq:massG0} is that $\dot{t}_{(0)}=0$ and therefore $t_{(0)}$ is constant and the evolution is in this sense instantaneous. Moreover, $\dot{E}_{(0)}=0$ and the pair of variables ($E_{(0)}, t_{(0)})$ is decoupled from the other variables.
For the other variables we find that $\vec{p}_{(0)}$ is a constant vector on the sphere with radius $k$ and we could take a Euclidean evolution by picking this space-like direction.

The next order action in $S= \sum_{n\geq 0} S_{(n)}$ with $S_{(n)}$ of order $\lambda^n$ then is
\begin{align}
\label{eq:massG1}
    S_{(1)} &= \lambda \int d\tau \bigg[ 
    -E_{(0)} \dot{t}_{(1)}  - E_{(1)} \dot{t}_{(0)} + \vec{p}_{(0)} \cdot \dot{\vec{x}}_{(1)} +\vec{p}_{(1)} \cdot \dot{\vec{x}}_{(0)} \nn\\
    &\hspace{20mm} - 
  \frac12 e_{(1)} \left( \vec{p}_{(0)}^{\, 2} - k^2 \right) - e_{(0)} \left( 2 \vec{p}_{(0)} \cdot \vec{p}_{(1)} - E_{(0)}^2 \right)
    \bigg]\,.
\end{align}
We note that the expansion of the momenta and coordinates always leads to a symplectic structure where the components are paired from opposite ends. The action enforces the constraints 
\begin{align}
    \phi_1 = \vec{p}_{(0)}^{\, 2} - k^2\,,
    \hspace{10mm}
    \phi_2 = 2 \vec{p}_{(0)} \cdot \vec{p}_{(1)} - E_{(0)}^2  
\end{align}
Moreover, the action~\eqref{eq:massG1} gives the constraint 
\begin{align}
\label{eq:dE10}
\dot{E}_{(1)}=0\,,
\end{align}
similar to the constraint $\dot{E}_{(0)}=0$ in the lowest order action~\eqref{eq:massG0}. 
An additional noteworthy point is that when starting from the phase space action~\eqref{eq:m0G} and expanding the phase space variables according to~\eqref{eq:expPS} one naturally ends up with what could be called an \textit{unconventional} symplectic structure where, at order $\lambda^N$, the variable $x^i_{(n)}$ is paired with $p^i_{(N-n)}$. This does not happen when starting from configuration space as in the previous section~\ref{sec:MGP} where the momentum $p^i_{(n)}$ was \textit{defined} as being conjugate to $x^i_{(n)}$.

The expanded actions $S_{(N)}$ derived from phase space always feature the same number $N+1$ of $t_{(n)}$ and $x^i_{(n)}$ by construction. This makes them a bit different from the configuration space actions such as~\eqref{eq:S2G}, where there is a superficial imbalance. As discussed at the end of the previous section, this imbalance can be rectified by introducing one more seemingly spurious variable into configuration space. This extra variable has a well-justified physical origin from Minkowski space. A similar phenomenon arises here since for all expanded $S_{(N)}$ from Galilean phase space the final pair $(E_{(N)}, t_{(0)})$ only enters the canonical action through its symplectic term and is completely decoupled from the rest. It carries no dynamics and leads to the very simple canonical constraint $\dot{E}_{(N)}=0$, see for instance~\eqref{eq:dE10}.

As we shall see later, there is an interesting connection of this constraint structure to that of particles in Carroll space-time and that we comment on in section~\ref{sec:GalCar}. The special role played by the final pair of canonical variables as well as the unconventional symplectic structure will be seen to enter in the connection.

An important observation here is that we have now transitioned to a phase space action. In the case of configuration space actions we could recover corrections to relativistic actions by combining a gauge choice with a choice of a slice condition, see for instance~\eqref{eq:GalGF} and~\eqref{eq:GalSC}. We are not aware of a similar construction for phase space.

\subsection{Particles in post-Carrollian space-time}
\label{sec:Cpart}

In this section we study the Carrollian limits of relativistic particles, using the post-Carrollian space-time introduced in section~\ref{sec:CarST}. 

\subsubsection{Massive Carroll particle}

In order to obtain the Carrollian expansion of a time-like Carroll
particle we will start by considering the canonical action of a 
time-like massive relativistic particle given by the Lagrangian
\begin{align}\label{relcanonical}
L_c=P_a \dot X^{a}-\frac e2(P^2+m^2c^2)\,.
\end{align}
We also use the Carrollian expansion of the collective coordinates (\ref{carrollexpansion})
\begin{align}
X^0 = \sum_{n\geq 0} s_{(n)} \lambda^{n+1/2} \,,\quad
X^i = \sum_{n\geq 0} x^i_{(n)} \lambda^{n}\,.
\end{align}
The first few terms of these expansions are explicitly
\begin{align}
X^0 =  \lambda^{1/2} s_{(0)}+\lambda^{3/2} s_{(1)}+\cdots\,,
\quad
X^i = x^i_{(0)} +\lambda x^i_{(1)}+\cdots
\end{align}
The expansion for the space-time momenta is given by
\begin{align}
P_0= -\sum_{n\geq 0} E_{(n)} \lambda^{n-1/2} \,,\quad
P^i = \sum_{n\geq 0} p^i_{(n)} \lambda^{n}\,.
\end{align}
In order to expand (\ref{relcanonical}) we also need the expansion of the einbein
\begin{align}
e = \sum_{n\geq 0} e_{(n)} \lambda^{n+1}
\end{align}
and also the rescaling (recall $\lambda=C^{-2}$)
\begin{align}
m c = M  \lambda^{-1/2}
\end{align}
that defines a new mass $M$.
The  relativistic action then  becomes
\begin{align}
L_c = \sum_{n\geq 0} L_{(n)} \quad\quad\text{with $L_{(n)}$ of order $\lambda^{n}$.}
\end{align}
The first terms of the expansion are
\begin{align}
 L_{(0)} =-E_{(0)}\dot s_{(0)}+\vec{p}_{(0)}\dot{\vec{x}}_{(0)}
 -\frac{e_{(0)}}2\left(-E_{(0)}^2+M^2\right)
\end{align}
which agrees with the one of \cite{Bergshoeff:2014jla,Duval:2014uoa}, and
\begin{align}
 L_{(1)} &=-E_{(1)}\dot s_{(0)}-E_{(0)}\dot s_{(1)}
 +\vec{p}_{(0)}\dot{\vec{x}}_{(1)}+\vec{p}_{(1)}\dot{\vec{x}}_{(0)}\nn\\
 &\hspace{10mm}
 -\frac{e_{(1)}}2\left(-E_{(0)}^2+M^2\right)-
 \frac{e_{(0)}}2\left(2E_{(0)}E_{(1)}-\vec{p}_{(0)}^{\,2}\right)\,.
\end{align}
This action derived from phase space also has the unconventional symplectic structure already encountered in section~\ref{sec:M0G}.
If we integrate out the auxiliary variables $(E_{(0)}, E_{(1)},$ $ e_{(0)}, e_{(1)})$ from this action we obtain
\begin{align}
L_{(1)} = \vec{p}_{(0)}\dot{\vec{x}}_{(1)}+\vec{p}_{(1)}\dot{\vec{x}}_{(0)}
 + M \dot{s}_{(1)}  + \frac{\dot{s}_{(0)}}{2M} \vec{p}_{(0)}^{\,2}\,.
\end{align}
The equations of motion obtained from this action by varying the $\vec{p}_{(i)}$ are
\begin{align}
\dot{\vec{x}}_{(0)} = 0 \,,\hspace{10mm} 
\dot{\vec{x}}_{(1)} = - \frac{\dot{s}_{(0)}}{M} \vec{p}_{(0)}  \,.
\end{align}
The $\vec{p}_{(i)}$ are constant and there are no further constraints by varying the $s_{(i)}$ whose values can be fixed by gauge invariance.  One important consequence now is that while the lowest order Carroll tachyon is well-known to be stationary at a fixed position, we now see that the correction in principle allows for a non-trivial motion. This is also what one would expect from a correction to the strict Carroll limit $c\to 0$ where the light-cone collapses to a line: The correction should open the light-cone slightly and thus allow for motion.

\subsubsection{Tachyonic Carroll particle}

In the strict Carroll limit, where the speed of light tends to zero, all moving particles have to be tachyonic as just argued. We therefore consider the action of a tachyonic particle that is constructed from the invariant metric
\begin{align}
ds^2 = \eta_{ab} dX^a dX^b = \sum_{m,n\geq 0} \lambda^{m+n} \left( - \lambda\, ds_{(m)} ds_{(n)} +  \delta_{ij} dx^i_{(m)} dx^j_{(n)} \right)\,,
\end{align}
where we used~\eqref{eq:MCinf}.
The tachyonic configuration space action to be expanded is then 
\begin{align}
\label{eq:Stach}
S =  M\lambda^{-1/2} \int d\tau \sqrt{\eta_{ab} \dot{X}^a \dot{X}^b} = S_{(0)} + S_{(1)} + S_{(2)} + \ldots\,,
\end{align}
where the difference to the massive Galilean particle~\eqref{eq:PAGal} is that the sign inside the square-root has changed since we are now dealing with a tachyon. Moreover, we express the mass of the tachyon as $mc = MC=\tilde{M}$ and recall that $\lambda=C^{-2}$. The `mass' $\tilde{M}$ does not have canonical dimensions, but this is compensated for by declaring $\tau$ to be of the Carroll time dimension $L^2/T$. 
The individual terms in the expanded action are then
\begin{align}
\label{eq:Ctach}
S_{(0)} &= \tilde{M} 
\int d\tau \sqrt{\delta_{ij} \dot{x}^i_{(0)} \dot{x}^j_{(0)}}\,,\nn\\
S_{(1)} &=    \tilde{M} C^{-2}
\int d\tau \frac{2\delta_{kl} \dot{x}^k_{(0)} \dot{x}^l_{(1)} - \dot{s}_{(0)}^2 }{2\sqrt{\delta_{ij} \dot{x}^i_{(0)} \dot{x}^j_{(0)}}}\,,\nn\\
S_{(2)} &=  \tilde{M} C^{-4} \int d\tau \Bigg[ \frac{\delta_{kl} \dot{x}_{(1)}^k \dot{x}_{(1)}^l + 2\delta_{kl} \dot{x}_{(0)}^k \dot{x}_{(2)}^l}{2 \sqrt{\delta_{ij} \dot{x}^i_{(0)} \dot{x}^j_{(0)}}} - \frac{\dot{s}_{(0)} \dot{s}_{(1)}}{ \sqrt{\delta_{ij} \dot{x}^i_{(0)} \dot{x}^j_{(0)}}}  \\
&\hspace{30mm} - \frac{\dot{s}_{(0)}^4}{8(\delta_{ij} \dot{x}^i_{(0)} \dot{x}^j_{(0)})^{3/2}} + \frac{\dot{s}_{(0)}^2 \delta_{kl} \dot{x}_{(0)}^k \dot{x}_{(1)}^l  - (\delta_{kl} \dot{x}_{(0)}^k \dot{x}_{(1)}^l)^2}{2(\delta_{ij} \dot{x}^i_{(0)} \dot{x}^j_{(0)})^{3/2}} \Bigg]\nn
\end{align}
and higher order terms can be obtained easily. The action $S_{(0)}$ has already been studied in~\cite{deBoer:2021jej}.

In order to elucidate the nature of these further actions, we now analyse them canonically.

\subsubsection*{Lowest order Carroll tachyon}

{}From the action $S_{(0)}$ in~\eqref{eq:Ctach} one finds the canonical momentum (using $\lambda^{-1/2}=C$)
\begin{equation}
p_i^{(0)} = \frac{\tilde{M} \dot{x}^i_{(0)}}{|\dot{\vec{x}}_{(0)}|}\,,
\end{equation}
where $|\dot{\vec{x}}_{(0)}| = \sqrt{\delta_{ij} \dot{x}^i_{(0)} \dot{x}^j_{(0)}}$. The canonical momentum therefore satisfies the primary (scalar) constraint
\begin{equation}
\label{eq:Tcon}
\phi_1 =  \frac12 \left(\delta^{ij} p_i^{(0)} p^{(0)}_j - \tilde{M}^2  \right)=0\,.
\end{equation}
This mass-shell constraint is first-class and generates the gauge transformations
\begin{equation}
\label{eq:GC1}
\delta x^i_{(0)} = \epsilon \delta^{ij} p_j^{(0)}\,,\quad
\delta p_i^{(0)} = 0
\end{equation}
in phase space. If one considers the action to formally also depend on the lowest order Carroll time $s_{(0)}$ we also get $E_{(0)} = 0$ as a constraint since the variable $\dot{s}_{(0)}$ does not appear in the action.
There are no further constraints.

The extended Hamiltonian action is
\begin{equation}
S_{(0)} =  \int d\tau \left[ p_i^{(0)} \dot{x}_{(0)}^i - e \phi_1 \right]=  \int d\tau \left[ p_i^{(0)} \dot{x}_{(0)}^i -\frac{e}{2} \left(\delta^{ij} p_i^{(0)} p^{(0)}_j - \tilde{M}^2\right)  \right]\,.
\end{equation}
The gauge symmetry~\eqref{eq:GC1} can be gauge-fixed by setting for instance the first spatial component $x^1_{(0)}= C^{-1} \tau$ (assuming $p_1^{(0)}\neq 0$ without loss of generality). The reduced phase space consists then of the transverse components $(x^\alpha_{(0)}, p_\alpha^{(0)})$ where $\alpha=2,3,\ldots, d$. The action on the reduced phase space is
\begin{equation}
S_{(0)} =  \int d\tau \left[ p_\alpha^{(0)} \dot{x}_{(0)}^\alpha \pm \sqrt{\tilde{M}^2 - \delta^{\alpha\beta} p_\alpha^{(0)} p_\beta^{(0)}}
\right]
\end{equation}
There is a choice of square root when solving the constraint $\phi_1=0$.
The Hamiltonian is no longer invariant under the full rotation group $SO(d)$ but only under an $SO(d-1)$ subgroup. Moreover, the energy is not bounded from below or above. A similar phenomenon has been observed for the Galilean string~\cite{Gomis:2016zur}.

\subsubsection*{First correction to Carroll tachyon}

{}From the action $S_{(1)}$ in~\eqref{eq:Ctach} we deduce the following conjugate momenta (setting $\tilde{M}=C=1$ for simplicity)
\begin{align}
p_i^{(0)} &= \frac{\dot{x}_{(1)}^i}{|\dot{\vec{x}}_{(0)}|}  - \frac{2 \dot{\vec{x}}_{(0)}\cdot \dot{\vec{x}}_{(1)} - \dot{s}_{(0)}^2}{2|\dot{\vec{x}}_{(0)}|} \dot{x}_{(0)}^i\,,\nn\\
p_i^{(1)} &= \frac{\dot{x}_{(0)}^i}{|\dot{\vec{x}}_{(0)}|}\,,\\
E^{(0)} &= -\frac{\delta S_{(1)}}{\delta \dot{s}_{(0)}} = \frac{\dot{s}_{(0)}}{|\dot{\vec{x}}_{(0)}|}\nn
\end{align}
and the two primary, first-class constraints
\begin{align}
\label{eq:ggC2}
\phi_1 &=  \delta_{ij} p^{(0)}_i p^{(1)}_j  - \frac12 E^{(0)}E^{(0)} =0\,,\nn\\
\phi_2 &= \frac12 \left(\delta^{ij} p_i^{(1)} p^{(1)}_j - 1 \right)=0\,.
\end{align}
Similar to the discussion at the end of section~\ref{sec:MGP} above, we could complement this by 
\begin{align}
E_{(1)}=0
\end{align}
 by thinking of the theory as depending on both space and (Carroll) time coordinates to second order by including $s_{(1)}$. There are no further constraints.
The first-class constraints~\eqref{eq:ggC2} generate the gauge transformations
\begin{align}
\delta_1 x^i_{(0)} &=  \epsilon_1 \delta^{ij} p_j^{(1)} \,,& \delta_1 p^{(0)}_i &=0\,,\nn\\
\delta_1 x^i_{(1)} &=  \epsilon_1 \delta^{ij} p_j^{(0)} \,,& \delta_1 p^{(1)}_i &=0\,,\nn\\
\delta_1 s_{(0)} &=  \epsilon_1 E^{(0)} \,,& \delta_1 E^{(0)} &=0
\end{align}
and
\begin{align}
\delta_2 x^i_{(0)} &=  0 \,,& \delta_2 p^{(0)}_i &=0\,,\nn\\
\delta_2 x^i_{(1)} &=  \epsilon_2 \delta^{ij} p_j^{(1)} \,,& \delta_2 p^{(1)}_i &=0\,,\nn\\
\delta_2 s_{(0)} &= 0\,,& \delta_2 E^{(0)} &=0\,.
\end{align}

The constraints can be gauge-fixed by setting $s_{(0)}=0$ (for $\phi_1$) and $x_{(1)}^1=C^{-1}\tau$ (for $\phi_2$). The reduced phase space then consists of $(x^i_{(0)}, p_{(0)}^i, x^\alpha_{(1)}, p_\alpha^{(1)})$ where $\alpha=2,3,\ldots, d$ labels the transverse coordinates. The Hamiltonian action on this reduced phase space is 
\begin{align}
S_{(1)} = \int d\tau \left[ p_i^{(0)} \dot{x}_{(0)}^i + p_\alpha^{(1)} \dot{x}_{(1)}^\alpha \pm \sqrt{1- \delta^{\alpha\beta}p_\alpha^{(1)} p_\beta^{(1)}} \right]\,.
\end{align}
The dynamics implied by this action is that $x_{(0)}^i=\text{const.}$ and $x_{(1)}^\alpha$ moves with the Euclidean time $\tau$ from the gauge-fixing. 

\subsubsection*{Configuration space actions and choice of slice}

As for the Galilean particle in~\eqref{eq:GalGF}, we can now consider a gauge-fixing in configuration space. 
Here, we use the freedom to think of the world-line parameter $\tau$ to be of the same dimension as Carroll time, meaning it has dimension $L^2/T$. Then the gauge choice we make is 
\begin{align}
 s_{(n)} = C^{2n} \tau\,.
\end{align}
Moreover, and similar to~\eqref{eq:GalSC}, we make the
 choice of slice 
\begin{align} 
x_{(n)}^i = C^{2n} x^i \,.
\end{align}
That this gauge choice is admissible can be checked using the gauge symmetries exhibited above.
Substituting these conditions into~\eqref{eq:Ctach} we obtain
\begin{align}
\tilde{S}_{(0)} &= \tilde{M} \int d\tau \, \sqrt{\dot{\vec{x}}^{\, 2}}\,,\nn\\
\tilde{S}_{(1)} &= \tilde{M} \int d\tau \left[ \sqrt{\dot{\vec{x}}^{\, 2}}- \frac{ 1}{2C^2\sqrt{\dot{\vec{x}}^{\, 2}}} \right]\,,\\
\tilde{S}_{(2)} &= \tilde{M} \int d\tau \left[ \sqrt{\dot{\vec{x}}^{\, 2}}- \frac{1}{2C^2\sqrt{\dot{\vec{x}}^{\, 2}}} - \frac{1}{8C^4 (\dot{\vec{x}}^{\,2})^{3/2}} \right]\,. \nn
\end{align}
This is to be compared to the large-$C$ expansion of the relativistic tachyon action~\eqref{eq:Stach}, now rewritten as ($x^0 = s/C$)
\begin{align}
S &= \tilde{M} \int d\tau \sqrt{ \dot{\vec{x}}^{\, 2} - \frac{\dot{s}^2}{C^2}}\nn\\
&=  \tilde{M} \int d \tau \left[ \sqrt{\dot{\vec{x}}^{\, 2}} - \frac{ \dot{s}^2 }{2C^2 \sqrt{\dot{\vec{x}}^{\, 2}}} - \frac{\dot{s}^4}{8  C^4(\dot{\vec{x}}^{\,2})^{3/2}} + \mathcal{O}(C^{-6})\right]\,,
\end{align}
whose gauge-fixed form with $s=  \tau$ agrees with the actions above.

\subsection{Relation between the expanded Galilean and Carrollian particle actions}
\label{sec:GalCar}

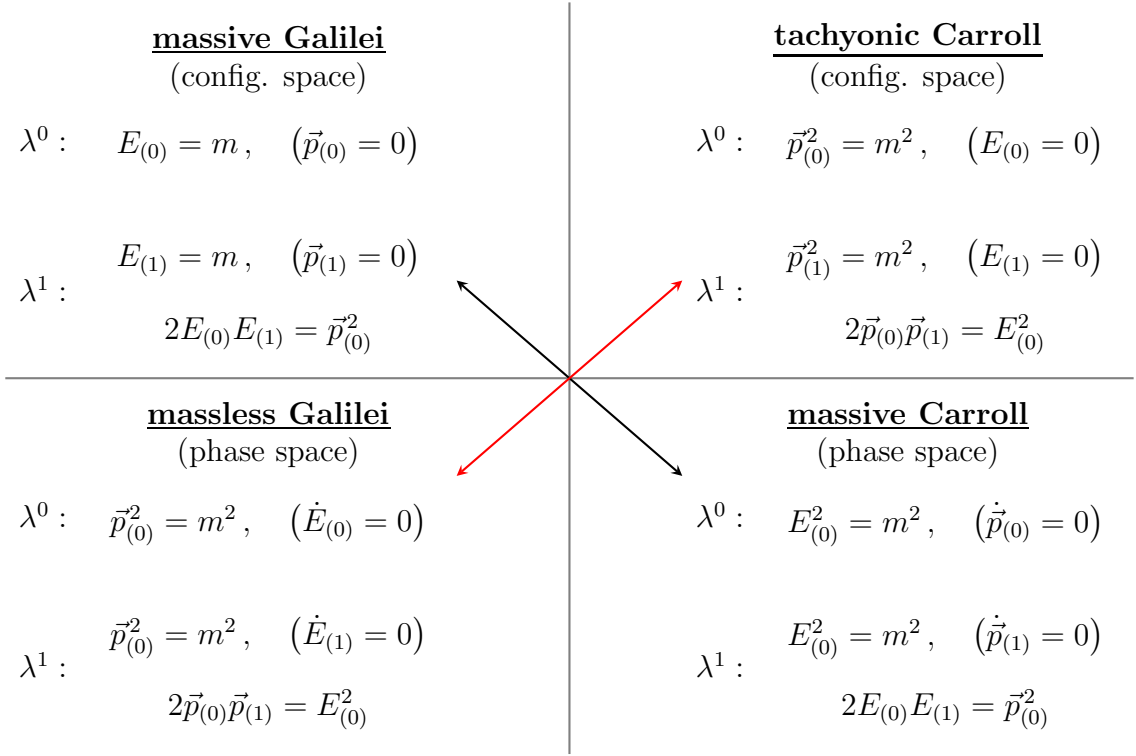
\begin{figure}[t!]
\begin{center}
\begin{tikzpicture}
\draw [thick,gray] (-7.5,0)--(7.5,0);
\draw [thick,gray] (0,-5)--(0,5);
\draw (-4,4.5) node{\underline{\textbf{massive Galilei}}};
\draw (-4,4) node{(config. space)};
\draw [anchor=north] (-7,3.5) node{$\lambda^0:$};
\draw [anchor=north] (-4,3.5) node{$E_{(0)}=m\,,\quad \big(\vec{p}_{(0)}=0\big)$};
\draw [anchor=north] (-7,1.5) node{$\lambda^1:$};
\draw [anchor=north] (-4,2) node{$E_{(1)}=m\,,\quad \big(\vec{p}_{(1)}=0\big)$};
\draw [anchor=north]  (-4,1) node{$2E_{(0)}E_{(1)}=\vec{p}_{(0)}^{\,2}$};
\draw (-4,-0.5) node{\underline{\textbf{massless Galilei}}};
\draw (-4,-1) node{(phase space)};
\draw [anchor=north] (-7,-1.5) node{$\lambda^0:$};
\draw [anchor=north] (-4,-1.5) node{$\vec{p}_{(0)}^{\,2}=m^2\,,\quad \big(\dot{E}_{(0)}=0\big)$};
\draw [anchor=north] (-7,-3.5) node{$\lambda^1:$};
\draw [anchor=north]  (-4,-3) node{$\vec{p}_{(0)}^{\,2}=m^2\,,\quad \big(\dot{E}_{(1)}=0\big)$};
\draw [anchor=north]  (-4,-4) node{$2\vec{p}_{(0)} \vec{p}_{(1)} = E_{(0)}^2$};
\draw (4.5,4.5) node{\underline{\textbf{tachyonic Carroll}}};
\draw (4.5,4) node{(config. space)};
\draw [anchor=north] (2,3.5) node{$\lambda^0:$};
\draw [anchor=north] (5,3.5) node{$\vec{p}_{(0)}^{\,2}=m^2\,,\quad \big(E_{(0)}=0\big)$};
\draw [anchor=north] (2,1.5) node{$\lambda^1:$};
\draw [anchor=north]  (5,2) node{$\vec{p}_{(1)}^{\,2}=m^2\,,\quad \big(E_{(1)}=0\big)$};
\draw [anchor=north]  (5,1) node{$2\vec{p}_{(0)} \vec{p}_{(1)} = E_{(0)}^2$};
\draw (4.5,-0.5) node{\underline{\textbf{massive Carroll}}};
\draw (4.5,-1) node{(phase space)};
\draw [anchor=north] (2,-1.5) node{$\lambda^0:$};
\draw [anchor=north] (5,-1.5) node{$E_{(0)}^2=m^2\,,\quad \big(\dot{\vec{p}}_{(0)}=0\big)$};
\draw [anchor=north] (2,-3.5) node{$\lambda^1:$};
\draw [anchor=north]  (5,-3) node{$E_{(0)}^2=m^2\,,\quad \big(\dot{\vec{p}}_{(1)}=0\big)$};
\draw [anchor=north]  (5,-4) node{$2E_{(0)}E_{(1)} = \vec{p}_{(0)}^{\,2}$};
\draw [thick,stealth-stealth] (-1.5,1.3)--(1.5,-1.3);
\draw [thick,red,stealth-stealth] (1.5,1.3)--(-1.5,-1.3);
\end{tikzpicture}
\end{center}
\caption{\label{fig:rel}\sl Diagram showing schematically the relation between the different non-relativistic limits for different types of particles. The focus is here on the constraints obeyed by the canonical variables. We have used the letter $m$ for all types of masses that appear. In the case of the massless Galilei particle this corresponds to the colour $k$, see~\eqref{eq:massG0}. The conditions shown in parentheses correspond to the ones that arise from using the same number of time and space variables as explained in the text.}
\end{figure}

There is a close relationship between the Galilean and Carrollian particle actions discussed in sections~\ref{sec:Gpart} and~\ref{sec:Cpart}. 
This can be seen by comparing for instance the constraints implied by the various actions in canonical form and the connection is illustrated in figure~\ref{fig:rel}. In the figure we have also illustrated whether or not the particle was obtained starting from phase space or configuration space in the preceding sections.

A special role is played by the conditions in parentheses. These arise as constraints for configuration space from considering the action to formally depend also on one more variable, namely $\vec{x}_{(N)}$ for massive Galilei and $s_{(N)}$ for tachyonic Carroll at order $\lambda^N$. These variables do not appear in the configuration space action  $S_{(N)}$ and therefore their conjugate momenta are constrained to vanish. Similarly, the expanded phase space actions contain the (unconventional) symplectic term $-\dot{t}_{(0)} E_{(N)}$ for massless Galilei and $\dot{\vec{x}}_{(0)} \cdot \vec{p}_{(N)}$ for massive Carroll with the canonical variables appearing nowhere else in the action and thus also decoupling completely. The associated equations of motion from phase space in parentheses  only imply constancy of a `top' canonical variable $\vec{p}_{(N)}$ or $E_{(N)}$. 
The variables appearing in the figure can also have different dimensions due to the different scalings used, but since we are exhibiting a similarity in structures we use the same letters in all cases. 

There is a known duality between Galilei and Carroll limits~\cite{Barducci:2019jhj} that acts horizontally in the diagram in each row. The duality was mentioned in the algebraic context in section~\ref{sec:contr} and it exchanges morally the spatial and temporal translations.\footnote{It is an exact duality in $1+1$ dimensions, in higher dimensions it is only heuristic since vector and scalar quantities are being interchanged.} This duality relates massive to tachyonic particles because of the interchange of the associate physical quantities $E\leftrightarrow \vec{p}$ in the mass-shell condition $E^2-\vec{p}^{\,2}=m^2$ which implies a change of sign of the squared mass.

On top of this, there is new relationship between Galilei and Carroll limits that acts across the diagonals, with only small differences.  
If one disregards the conditions in parentheses one can construct maps between the other constraints across diagonals as follows. At order $\lambda^N$ for the NW-SE diagonal (black arrow) one exchanges $E_{(n)}\leftrightarrow E_{(N-n)}$ and $p_{(n)}\leftrightarrow p_{(N-1-n)}$. The reason for treating the $E_{(n)}$ and $p_{(n)}$ slightly differently is due to the fact that $p_{(N)}$ appears in the special condition in parentheses. Since we only allow for positive energies in the massive Galilei case by construction, the constraints there contain a choice of square of the constraints in the massive Carroll case.

Similarly, the SW-NE diagonal (red arrow) corresponds to the map $E_{(n)}\leftrightarrow E_{(N-1-n)}$ and $p_{(n)}\leftrightarrow p_{(N-n)}$ where now $E_{(N)}$ is treated in a special way since it enters the special constraints.
The special condition in parentheses at lowest order is related to the energy of the particle. The zero energy condition was important in recent cosmological applications~\cite{deBoer:2021jej}. For the massless Galilei the condition follows from the equation of motion only requires the energy to be a constant but does not determine this constant and is thus weaker.
 Intriguingly, the dynamics in the reduced phase space is identical in both cases.\footnote{The massless Galilean particle has also appeared in the context of the optical Hall effect~\cite{Duval:2005ry} where the appropriate Galilean coadjoint orbits were used.}

We have verified explicitly that the maps indicated above also hold at the next order in $\lambda$ and from the construction of the actions it seems clear that this correspondence will to any order.

We note that a feature of both types of dualities (horizontal and across the diagonal) is that the number of degrees of freedom is not preserved.
As an example in $D=3+1$, the order $\lambda^0$ of massive Galilei has no degrees of freedom as there are four first-class constraints for eight phase space variables. By contrast, the horizontally mapped tachyonic Carroll has only two first-class constraints for eight variables and therefore four degrees of freedom in phase space, corresponding to the direction of the motion of the tachyon. 
Going across the diagonal to massive Carroll at order $\lambda^0$, one finds six degrees of freedom in phase space that correspond the arbitrary position of the Carroll particle and the components of $\vec{p}$ (which are unrelated to velocity).

\subsection{Particle in conformal Galilean space-time}
\label{sec:GCApart}

Massless relativistic particles in Minkowski space enjoy more symmetries than massive ones in that the global symmetry is extended from the Poincar\'e algebra to the conformal algebra. This can be seen by looking at the action
\begin{align}
S = \int d\tau \frac{\dot{X}^a \dot{X}_a}{2e}
\end{align}
and checking invariance of the equations under the transformations~\eqref{eq:confX}. For this one has to also consider the relativistic transformation $\delta_{\beta^a S_a} e = 4 \beta_a x^a e$  of the einbein  under special conformal transformations. Under dilatations, the einbein scales as $\delta_{\sigma D} e = 2 \sigma e$.

Expanding this action expressed in terms of the collective coordinates~\eqref{eq:Xconf} leads to $S= S_{(0)}+ S_{(1)} + \ldots$ with
\begin{align}
\label{eq:zmG}
S_{(0)} &=   \int d\tau  \frac{-\dot{t}_{(0)}^2}{2e_{(0)}}   \,,\nn\\
S_{(1)} &= \lambda  \int d\tau  \frac{- 2 e_{(0)} \dot{t}_{(0)} \dot{t}_{(1)} + \delta_{ij} e_{(0)} \dot{x}^i_{(0)} \dot{x}^j_{(0)} + e_{(1)} \dot{t}_{(0)}^2}{2e_{(0)}^2}
\end{align}
where we have also expanded the einbein according to $e=\sum_{n\geq 0} e_{(n)} \lambda^n$.  For the expanded components this implies the following transformations under special conformal transformations
\begin{align}
\delta e_{(0)} = - 4 b_{(0)} t_{(0)} e_{(0)} \,, \quad
\delta e_{(1)} = -4 b_{(1)} t_{(0)} e_{(0)} - 4 b_{(0)} t_{(1)} e_{(0)} -4 b_{(0)} t_{(0)} e_{(1)} + 4\delta_{ij} \beta_{(0)}^i x_{(0)}^j e_{(0)}\,.
\end{align}
One can verify that these transformations, together with~\eqref{eq:expconf}, leave the actions $S_{(0)}$ and $S_{(1)}$ invariant.

In order to see the physical degrees of freedom of 
$S_{(0)}$ and $S_{(1)}$ we do the Hamiltonian analysis.
The momenta are
\begin{align}
\pi_{ e_{(0)}} = 0\,, \quad
E_{(0)}= -\frac{\partial L_{(0)}}{\partial\dot t_{(0)}}=
\frac{\dot t_{(0)}}{e_{(0)}}
\end{align}
and the canonical Hamiltonian is 
\begin{align}
H_c = \frac{{e_{(0)}}}{2}E_{(0)}^2
\,.
\end{align}
We  have two first-class constraints $\pi_{ e_{(0)}} = 0\,,
E_{(0)}^2=0$; the second constraint is irregular and we should consider $E_{(0)}=0$ as an effective constraint, see for example \cite{Miskovic:2003ex}.
Since the dimension of the phase space is four and we have two first-class constraints there no physical degrees of freedom.
In the case of $S_{(1)}$ we have four constraints
\begin{align}
\pi_{ e_{(0)}} &= 0\,, & \pi_{ e_{(1)}} &= 0\,, \nn\\
  \phi_{(0)} & = 2 E_{(0)}E_{(1)}- \vec p_{(0)}^{\, 2}
  \,, & E_{(1)}^2&=0\,.
\end{align}
Again, there are irregular constraints the true effective constraints
are 
\begin{align}
\pi_{ e_{(0)}} = 0\,, \quad\quad \pi_{ e_{(1)}} = 0 \,, \quad\quad
\vec p_{(0)}=0\,,\quad\quad E_{(1)} = 0
\,.
\end{align}
The number of physical degrees of freedom is different from zero this time, and we are left with two degrees in phase space, namely $t_{(0)}$ and $E_{(0)}$. The fact that the number of degrees of freedom changes with the order in the expansion also occurs in the other cases prior to the choice of a slice. Once a slice condition is applied the number of degrees of freedom is unchanged.

\subsection{Particle in curved background}
\label{sec:AdSpart}

We now study massive particle dynamics that are invariant under the extended algebra with generators~\eqref{eq:AdSexp}. As shown in~\cite{Gomis:2020wrv}, expanding the usual (A)dS invariant particle metric using the coordinates~\eqref{eq:AdSco} can be done in a way similar to non-relativistic cases and leads at lowest orders to the following actions
\begin{align}
S_{(0)} &= \frac{m}{2} \int  d\tau \, \dot{x}_{(0)}^2\,,\nn\\
S_{(1)} &= \frac{m}{2} \int d\tau \left[ \dot{x}_{(0)} {\cdot} \dot{x}_{(1)} +\frac{\sigma}{6}\left( x_{(0)}^2 \, \dot{x}_{(0)}^2  - (x_{(0)}{\cdot} \dot{x}_{(0)})^2 \right) \right]\,, \text{etc.}
\end{align}
where $m$ is the mass of the particle and all contractions are done with the Minkowski metric $\eta_{ab}$. The 

Putting now together the equation of motion for the collective coordinate
\begin{align}
x^a = \sum_{n>0} R^{-2m-1} x_{(m)}^a 
\end{align}
we find from the individual equations of motion (when evaluated at a given fixed order) that
\begin{align}
\ddot{x}^a &= \frac{2\sigma}{3R^2} \left(\dot{x}^2 x^a - x{\cdot}\dot{x} \dot{x}^a \right) + \frac{2}{45R^4}\left( x^2 (x{\cdot}\dot{x}) \dot{x}^a +3 x^2 \dot{x}^2 x^a - 4(x{\cdot}\dot{x})^2 x^a \right) + \ldots
\end{align}
This equation can be checked to agree with the expansion of the geodesic equation of a massive particle on an (A)dS background for large radius of curvature $R$~\cite{Gomis:2020wrv}, written in appropriate coordinates where the metric takes the form
\begin{align}
ds^2 = dx^a \eta_{ab} dx^b + \left( \frac{\sinh^2 r}{r}-1\right) dx^a \mathcal{P}_{ab} dx^b
\end{align}
where $r= \sigma x^a\eta_{ab} x^b$ and $\mathcal{P}_{ab} = \eta_{ab} -\frac{x_a x_b}{x^2}$. 
Therefore, we conclude again that the infinite expansion of the symmetry allows us to recover the expansion of dynamics in the desired limit of small curvature. 

\subsection{Particle in electro-magnetic background}
\label{sec:Maxpart}

Particles in electric-magnetic backgrounds are subject to the Lorentz force, where the relativistic equation of motion can be written as
\begin{align}
\label{eq:Lorentz}
m\ddot{x}_a =  F_{ab} \dot{x}^b\,,
\end{align}
where we have set the electric charge of the massive particle to one. When the electro-magnetic field is constant, the Poincar\'e symmetry is broken to translations $P_a$ and $F^{ab} M_{ab}$ as well as $\varepsilon^{abcd} F_{ab}M_{cd}$ in $D=4$~\cite{Bacry:1968zf}. If one considers the space of all constant electro-magnetic fields $F_{ab}$ with the obvious action of the Lorentz algebra one can maintain the whole Poincar\'e algebra. One can also consider constant shifts of $F_{ab}$ by introducing a new generator $Z_{ab}$ and the resulting system then is invariant under the Maxwell algebra where $Z_{ab}= [P_a,P_b]$~\cite{Schrader:1972zd}. 

In order to describe varying electro-magnetic fields one has to consider an even further extension of the Maxwell algebra as shown originally in~\cite{Bonanos:2008ez}. Here, we recall how this works in a free Lie algebra language~\cite{Gomis:2017cmt}, where we use the free Lie algebra discussed in section~\ref{sec:Maxfree}.

The starting point is a non-linear realisation of the Maxwell free Lie algebra where the local symmetry is just the Lorentz symmetry. This means that we are considering a coset element whose gauge-fixed form is
\begin{align}
g = e^{x^a P_a} e^{\frac12 \theta^{ab} Z_{ab}} e^{\frac12 \xi^{ab,c} Y_{ab,c}} \cdots\,,
\end{align}
using the coordinates introduced in~\eqref{eq:Maxco}. The corresponding Maurer--Cartan form
\begin{align}
\label{eq:MCMax}
\Omega &= g^{-1} dg = dx^a P_a + \frac12 \left( d\theta^{ab} +dx^a x^b\right) Z_{ab} + \frac12\left( d\xi^{ab,c} -\theta^{ab} dx^c+\frac13 dx^a x^b x^c\right) Y_{ab,c} + \ldots \nn\\
&= \sum_{\ell=1}^\infty \Omega_{(\ell)}
\end{align}
has an expansion in terms of the levels of the free Lie algebra generated by the $P_a$. 

We then consider the particle action given by the Lagrangian
\begin{align}
\label{eq:LMax}
L d\tau = m\sqrt{-\Omega_a \Omega^a}  + \frac12 f_{ab} \Omega^{ab} + \frac12 f_{ab,c} \Omega^{ab,c} + \ldots
\end{align}
where the various $\Omega^a$, $\Omega^{ab}$, $\Omega^{ab,c}$ are the pull-backs of the components of the Maurer--Cartan form~\eqref{eq:MCMax} in an obvious way. The fields $f_{ab}$, $f_{ab,c}$ are new dynamical quantities whose transformation under the Lorentz symmetry is dual to that of the components of the Maurer--Cartan form. Note that the first component $\Omega^a$ has been treated differently, namely in such a way that it would just give a free massive Poincar\'e particle. 

The equations of motion implied by~\eqref{eq:LMax} are such that one always has~\cite{Gomis:2017cmt}
\begin{align}
m \ddot{x}_a = f_{ab} \dot{x}^b\,,
\end{align}
resembling the Lorentz force equation. While this equation is universal, the dynamical field $f_{ab}$ is obeying its own equation of motion that needs to be solved. However, the Lagrangian also implies equations for the other fields that, in the truncation to level $\ell\leq 3$ as shown in~\eqref{eq:LMax}, lead to
\begin{align}
\dot{f}_{ab,c}  &= 0\,, &
\dot{f}_{ab} &= -f_{ab,c} \dot{x}^c\,,\nn\\
\dot{\xi}^{ab,c} &= \frac13\left(2\theta^{ab}\dot{x}^c - \theta^{ca}\dot{x}^b - \theta^{bc} \dot{x}^a\right) -\frac16\left( \dot{x}^a x^b x^c - \dot{x}^b x^c x^a\right)\,,& \dot{\theta}^{ab} &= -\frac12\left(\dot{x}^a x^b - \dot{x}^b x^a\right)\,.
\end{align}
The evolution of the extra coordinates $\theta^{ab}$ and $\xi^{ab,c}$ is therefore determined\footnote{up to integration constants that reflect the global Maxwell symmetry} by that of the lowest coordinate $x^a$. The form of this dependence resembles a multipole expansion of a system of particles~\cite{Dixon:1967}. By contrast, the first line introduces other integration constants for the dynamical $f$-fields. In the truncation shown we can solve the corresponding equations and arrive at
\begin{align}
\label{eq:EM}
f_{ab} = f_{ab}^0 + f_{ab,c}^0 x^c + \ldots\,,
\end{align}
where the superscript ${}^0$ indicates an integration constant. In the previous equation we recognise the beginning of a Taylor expansion of an electro-magnetic field in Minkowski coordinates. Therefore, the extended Maxwell space-time has the potential to accommodate arbitrary electro-magnetic fields. 

This can be made more precise by considering the next level in the expansion~\cite{Bonanos:2008ez,Gomis:2017cmt}. This reveals that the full free Maxwell Lie algebra has too many generators compared to the Taylor expansion. In particular, there are generators that result in non-integrable contributions to $f_{ab}$, meaning that the field does not satisfy the Bianchi identity $\partial_{[a} f_{bc]} = 0$. To the level shown in~\eqref{eq:EM} this is guaranteed by the Young symmetry~\eqref{eq:Young3} but it fails in general. 

One can guarantee integrable field strengths by restricting the Maxwell free Lie algebra consistently to a quotient, namely the derivative quotient shown in~\eqref{eq:derivq}~\cite{Gomis:2017cmt}. We note that this kind of expansion is similar to what arises in unfolded dynamics~\cite{Vasiliev:2005zu,Boulanger:2015mka}. 
An open problem is  the precise connection of the behaviour of the higher coordinates $\theta^{ab}$, $\xi^{ab,c}$ and so on to multipole moments~\cite{Dixon:1967}. 
Moreover, in the analysis above the electro-magnetic field was a background field and it would be interesting to extend the analysis such that it becomes dynamical, i.e., such that the Maxwell equations also emerge.

\section{Conclusions}
\label{sec:concl}

In this paper, we have studied the algebraic structures of corrections to kinematic algebras, using the methods of Lie algebra expansions and free Lie algebras. This has allowed us to describe several physically interesting situations starting from generalised configuration spaces and by considering particle actions associated with them. From these we could recover systematically corrections to strict (non-relativistic, flat space, field free) limits. We paid particular to attention Carroll limits and their relation to Galilei. 
It would be interesting to exploit the Galilei/Carroll dualities and relations put forward in section~\ref{sec:GalCar} for applications such as gravity or hydrodynamics.

There are several avenues opened up by our approach. The first one is to extend our construction of generalised configuration spaces to that of generalised phase spaces and to see which conditions are needed to recover systematically corrections in phase space language. Besides particle actions one could also consider extended objects as probes. There are typically many more kinematic set-ups available due to the extended nature of the object~\cite{Brugues:2004an,Andringa:2012uz,Batlle:2016iel,Barducci:2019jhj,Bergshoeff:2020xhv}. We anticipate a similar multitude of generalised configuration spaces.

The particle actions considered in this paper were obtained either geometrically, using the invariant metrics of the expanded algebras, or from their corresponding phase space versions. An alternative approach to particle actions is given by non-linear realisations~\cite{Coleman:1969sm,Callan:1969sn,Salam:1969rq,Isham:1971dv,Volkov:1973vd,Ogievetsky:1974} whose generalisation to our infinite-dimensional algebras would be interesting to explore in detail. Non-linear realisations are also tied closely to the method of co-adjoint orbits that have not been studied for expanded algebras to the best of our knowledge.\footnote{For the case of affine algebras, studies of co-adjoint orbits can be found for example in~\cite{Reyman:1979,Khesin:2009}.}

 Another interesting possibility to explore could be the possible interaction among tachyonic Carroll particles.
Let us first consider two free tachyonic Carroll particles with spatial positions $\vec{x}_1$ and $\vec{x}_2$ and whose action is given by
\begin{align}
\label{eq:Ctachfree}
S=S^1_{(0)} +S^2_{(0)}&=  
\int d\tau \left[\sqrt{\tilde{M}_1^2 \, \dot{\vec{x}}_1\cdot\dot{\vec{x}}_1}+
 \sqrt{\tilde{M}_2^2\, \dot{\vec{x}}_2\cdot \dot{\vec{x}}_2}\right]
\end{align}
where $\tilde{M}_i$ are the masses of the two particles.
Like in a model of two interacting relativistic particles \cite{Kamimura:1977dv,Dominici:1977fh} we introduce the interaction among them by considering masses that depend on the relative position $r=\sqrt{(\vec{x}_1-\vec{x}_2)^2}$ of the coordinates 
\begin{align}
\tilde{M}_i^2(r)=\tilde{M}_i^2-V(r)\,,
\end{align}
where $V$ is a scalar function under spatial rotations.
The action for the interacting model is given by~\eqref{eq:Ctachfree} with the substitution $\tilde{M}_i^2 \to \tilde{M}^2_i(r)$.
The primary constraints of the model are 
\begin{align}
\tilde{\phi}_i = \vec{p}_i^{\, 2}-\tilde{M}_i^2(r)=0\,.
\end{align}
There is also the secondary constraint 
$(\vec{p}_1+\vec{p}_2) \cdot(\vec{x}_1-\vec{x}_2)=0$. 
 The possible physical implications of this model will be analysed elsewhere.

Our analysis was restricted to particle models and it would be interesting to generalise it to field theory. A bridge in that direction might be provided by world-line descriptions of field theory processes, see for instance~\cite{Feynman:1950ir,Casalbuoni:1974pj,Schubert:1996jj}. Among other things this requires a quantisation and generalisation of our considerations to interacting systems.
Different non-relativistic limits of field theories can be studied by considering limits of the ratio between the `electric' and `magnetic' contributions to a field's Hamiltonian energy, see for instance~\cite{Henneaux:2021yzg}. The electric contribution is the one due to time derivatives of the field while the magnetic one stems from space derivatives. As these two are related by the speed of light, making one larger than the other can also be thought of as a limit in the speed of light and therefore directly suggests to identify the electric limit as the Carroll limit and the magnetic limit as the Galilei limit. Whether this intuitive picture holds up to a more detailed study when applying the world-line picture to field theory is left to future work.

\section*{Acknowledgments}

We thank Jakob Palmkvist, Diederik Roest and Patricio Salgado-Rebolledo for very enjoyable collaborations that underlie some of the results presented here.
We are grateful to E.~A.~Bergshoeff, R.~Casalbuoni, J.~Figueroa-O'Farrill,
H.~Godazgar, M.~Godazgar, M.~Henneaux, C.~N.~Pope and P.~K.~Townsend for discussions.
The work of JG has been supported in part by MINECO FPA2016-76005-C2-1-P 
and PID2019-105614GB-C21 and from the State Agen\-cy for Research of the
Spanish Ministry of Science and Innovation through the Unit of Excellence
Maria de Maeztu 2020-203 award to the Institute of Cosmos Sciences
(CEX2019-000918-M).


\providecommand{\href}[2]{#2}\begingroup\raggedright\endgroup

\end{document}